\documentclass[twocolumn,nofootinbib,preprintnumbers,floatfix,aps,prl,10pt]{revtex4-2}
\usepackage{amsmath,amssymb,graphicx,booktabs,bm,psfrag,color,slashed,euscript,mathtools}
\usepackage{hyperref}
\usepackage{lipsum} 
\usepackage{bbold}
\usepackage{tikz}
\usepackage[export]{adjustbox}
\newcommand{\spac}{{\hspace{0.3mm}}}

\usepackage[T1]{fontenc}

\makeatletter 
    
\renewcommand\onecolumngrid{
\do@columngrid{one}{\@ne}%
\def\set@footnotewidth{\onecolumngrid}
\def\footnoterule{\kern-6pt\hrule width 1.5in\kern6pt}%
}

\renewcommand\twocolumngrid{
        \def\footnoterule{
        \dimen@\skip\footins\divide\dimen@\thr@@
        \kern-\dimen@\hrule width.5in\kern\dimen@}
        \do@columngrid{mlt}{\tw@}
}%

\makeatother    

\begin{document}
\preprint{MITP-23-033}
\title{Resummation of Next-to-Leading Non-Global Logarithms at the LHC}

\author{Thomas Becher$^a$}
\author{Nicolas Schalch$^a$}
\author{Xiaofeng Xu$^b$}
\affiliation{%
$^a$Institut f\"ur Theoretische Physik {\em \&} AEC, Universit\"at Bern, Sidlerstrasse 5, CH-3012 Bern, Switzerland\\
$^b$PRISMA$^+$ Cluster of Excellence, Johannes Gutenberg University, 55099 Mainz, Germany
}%

\begin{abstract}
In cross sections with angular cuts, an intricate pattern of enhanced higher-order corrections known as non-global logarithms arises. The leading logarithmic terms were computed numerically two decades ago, but the resummation of subleading non-global logarithms remained a challenge that we solve in this Letter using renormalization group methods in effective field theory. To achieve next-to-leading logarithmic accuracy, we implement the two-loop anomalous dimension governing the resummation of non-global logarithms into a large-$N_c$ parton shower framework, together with one-loop matching corrections. As a first application, we study the interjet energy flow in $e^+e^-$ annihilation into two jets. We then present, for the first time, resummed predictions at next-to-leading logarithmic accuracy for a gap-between-jets observable at hadron colliders.
\end{abstract}

\maketitle

 {\em Introduction. --- \/} There has been impressive progress in the perturbative calculation of processes at the Large~Hadron~Collider~(LHC). However, for observables involving disparate scales, computations beyond fixed perturbative order are necessary. These include cross sections involving a hard scale $Q$ but with sensitivity to a soft scale $Q_0$. Such cross sections involve large logarithms in the scale ratio $L=\ln ({Q}/{Q_0})$ that degrade the perturbative expansion and should be resummed to all orders to obtain reliable predictions. 

A generic set of observables involving scale hierarchies are cross sections where hard radiation is vetoed in certain angular regions. Prime examples are exclusive jet cross sections which require a veto on additional hard jets. While ubiquitous, the all-order resummation of such observables is challenging,  since they involve a complicated pattern of enhanced higher-order corrections known as non-global logarithms, which arises due to secondary emissions off hard partons~\cite{Dasgupta:2001sh,Dasgupta:2002bw,Banfi:2002hw}. At leading-logarithmic (LL)~$\sim \left(\alpha_s L \right)^n$ accuracy, resummed results both at large~\cite{Dasgupta:2001sh,Dasgupta:2002bw,Banfi:2002hw} and finite~$N_c$~\cite{Hatta:2013iba,Hagiwara:2015bia,Hatta:2020wre,DeAngelis:2020rvq} are available. Despite continued progress in the understanding of non-global observables over the past 20 years \cite{Weigert:2003mm,Forshaw:2006fk,Forshaw:2009fz,DuranDelgado:2011tp,Schwartz:2014wha,Becher:2015hka,Becher:2016mmh,Larkoski:2015zka,CaronHuot:2015bja,Becher:2016omr,Neill:2016stq,Caron-Huot:2016tzz,Larkoski:2016zzc,Hatta:2017fwr,Becher:2017nof,AngelesMartinez:2018cfz,Neill:2018yet,Balsiger:2018ezi,Dreyer:2018nbf,Balsiger:2019tne,Balsiger:2020ogy,Dasgupta:2020fwr,Hamilton:2020rcu,Caletti:2021oor,Becher:2021zkk,vanBeekveld:2022zhl,vanBeekveld:2022ukn,vanBeekveld:2023lfu}, a full resummation of next-to-leading logarithmic (NLL)~$\sim \alpha_s \left(\alpha_s L \right)^n$ corrections  remained elusive. 
 In this Letter we solve this problem based on a factorization theorem \cite{Becher:2015hka,Becher:2016mmh} obtained in soft-collinear effective field theory \cite{Bauer:2001yt,Bauer:2002nz,Beneke:2002ph}. 
The factorization theorem splits the cross section into hard and soft functions. To resum the large logarithms, one solves the renormalization group (RG) equations of the hard functions to evolve them from a scale $\mu \sim Q$ down to $\mu \sim Q_0$. Since the associated anomalous dimension is a matrix in the (infinite) space of  particle multiplicities, we resort to Monte Carlo (MC) methods to solve the RG equations. A key ingredient for NLL resummation is the recently extracted two-loop anomalous dimension \cite{Becher:2021urs} that we implement into a parton shower framework, which iteratively generates additional emissions to solve the RG equations. Combined with the one-loop corrections to the hard and soft functions we obtain in this Letter the full set of NLL contributions for gap-between-jets cross sections at lepton and hadron colliders. For the lepton-collider case NLL results were first presented in \cite{Banfi:2021xzn},  based on a very different formalism~\cite{Banfi:2021owj}, and we find full agreement within numerical uncertainties.
\\[12pt]
\begin{figure}
\includegraphics[scale=0.55]{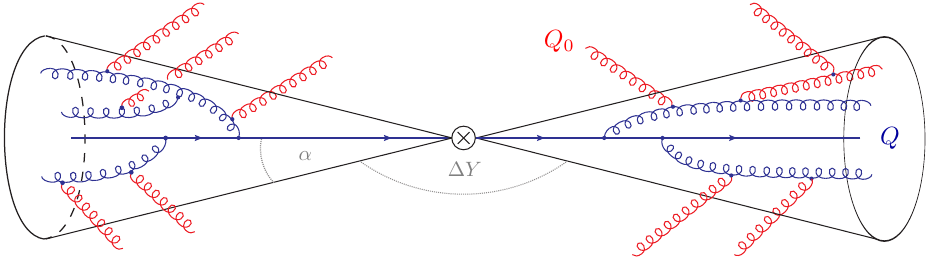}
\caption{Representation of the factorization formula \eqref{eq:factthm}. The blue lines depict hard radiation associated with the energy scale $Q$, which is constrained inside the jet cones, while the red lines represent the soft radiation at lower energies $Q_0$. The soft radiation can cover the entire phase space.
}\label{fig:interjet}
\end{figure}
 {\em Methodology. --- \/} The basis for our resummation are factorization theorems for jet production in the presence of a veto on radiation in certain angular regions of the phase space.  The simplest case is two-jet production in $e^+e^-$ collisions, which factorizes as \cite{Becher:2015hka,Becher:2016mmh}
\begin{align}
\hspace*{-0.23cm}\sigma(Q,Q_0) = \sum_{m=2}^{\infty} \big\langle \bm{\mathcal{H}}_m(\{\underline{n}\},Q,\mu) \otimes \bm{\mathcal{S}}_m(\{\underline{n}\},Q_0,\mu)\big\rangle\,,\label{eq:factthm}
\end{align}
 where $Q$ is the center-of-mass energy and $Q_0$ is the energy scale above which we veto radiation in the gap outside the jet cones. We impose the veto by demanding that the transverse energy $E_T$ of the particles in the gap is below $Q_0$. At the order we are working, our constraint is equivalent to imposing that the transverse momentum of the leading jet in the gap region is below $Q_0$. Figure \ref{fig:interjet} shows a pictorial representation of the factorization theorem \eqref{eq:factthm}. The hard functions $\bm{\mathcal{H}}_m$ describe $m$ hard partons, which we treat as massless, inside the jet cones. To obtain $\bm{\mathcal{H}}_m$, one integrates the squared amplitudes over the energies of the $m$ hard partons while keeping their directions $\{\underline{n}\}=\{n_1, \dots,n_m\}$ fixed. The bare hard functions in $d=4-2\epsilon$ are defined as  
\begin{align}
   \hspace*{-1mm}\bm{\mathcal{H}}_m 
   = & \hspace*{1mm} \frac{1}{2\spac Q^2} \prod_{i=1}^m \int\frac{dE_i\,E_i^{d-3}}{{\tilde{c}}^\epsilon\,(2\pi)^{2}}\,
    |\mathcal{M}_m(\{\underline{p}\})\rangle \langle\mathcal{M}_m(\{\underline{p}\})| \nonumber
    \\
   &\times (2\pi)^d\,\delta\Big( Q-\sum_{i=1}^m E_{i}\Big)\,
    \delta^{(d-1)}(\vec{p}_{\rm tot})\,\Theta_{\rm in}\!
    \left(\left\{\underline{n}\right\}\right) ,\label{eq:hardfct}
\end{align}%
with $\vec{p}_{\rm tot}$ the total momentum of the final state particles. The constraint $\Theta_{\rm in}\left(\left\{\underline{n}\right\}\right)=\theta_{\rm in}(n_1)\,\theta_{\rm in}(n_2)\, \dots \,\theta_{\rm in}(n_m)$ prevents the hard radiation from entering the veto region, i.e., it forces the hard partons to be inside the jet region. The constant $\tilde{c} = e^{\gamma_E}/\pi$ was introduced in \cite{Becher:2021urs}. 
\begin{figure}
\includegraphics[scale=0.75,valign=t]{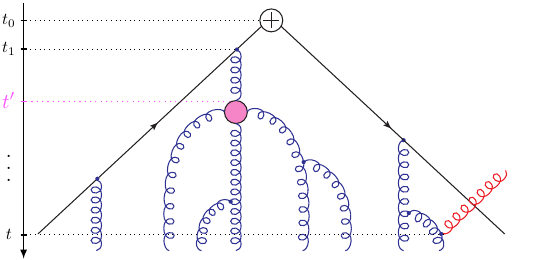}
\caption{\label{fig:shower} Pictorial representation of the NLL shower. Black lines represent the initial hard partons in $\bm{\mathcal{H}}_2(\mu_h)$. Blue lines denote hard emissions inside the jets generated by the shower evolution. The red line depicts a soft emission into the veto region, which terminates the shower. The pink blob is an insertion of the two-loop double-real contribution $\bm{d}_m$ at time~$t'$. } 
\end{figure}
Each hard particle may emit soft radiation, which we describe through a Wilson line $\bm{S}_i(n_i)$ along the corresponding direction $n_i$. The soft function $\bm{\mathcal{S}}_m$ is defined as the matrix element of these Wilson lines
\begin{align}
\bm{\mathcal{S}}_m 
    = & \hspace*{1mm} \int\limits_{X_s}\hspace{-0.52cm}\sum \,\langle  0 |\, \bm{S}_1^\dagger(n_1) \,  \dots\,  {\bm S}_m^\dagger(n_m)\,  |X_s \rangle\langle  X_s | \,\bm{S}_1(n_1) \,  \dots\, \nonumber \\
    & \hspace{1.3cm} \times \dots\,  {\bm S}_m(n_m) \, |0 \rangle \, \theta( Q_0 - E_{T,{\rm \, out}}),
\end{align}
where the constraint is imposed on the transverse energy in the veto region outside the jets. Since both hard and soft functions are matrices in the color space \cite{Catani:1996vz} of the $m$ partons, we take the color trace, as indicated by~$\langle \spac \cdots \rangle$. Eventually, in order to obtain the cross section, we integrate over all directions $\left\{\underline{n}\right\}$ which is indicated by the symbol $\otimes$. 

The bare hard and soft functions suffer from divergences that can be removed by renormalization. The cross section \eqref{eq:factthm} must be independent of the associated renormalization scale $\mu$ which leads to RG equations
\begin{align}\label{eq:hrdRG}
\mu \frac{\mathrm{d}}{\mathrm{d}\mu}\,\bm{\mathcal{H}}_m  &= - \sum_{l =2}^{m}  \bm{\mathcal{H}}_l \, \bm{\Gamma}_{lm} \,.
\end{align}
To carry out the resummation of large logarithms, we compute the functions $\bm{\mathcal{H}}_m$ at the hard scale $\mu_h\sim Q$ and use the RG equations \eqref{eq:hrdRG} to evolve them to the soft scale $\mu_s\sim Q_0$, where we evaluate the functions $\bm{\mathcal{S}}_m$. After integrating over the directions $\left\{\underline{n}\right\}$ of the hard partons, we obtain a resummed prediction for the cross section. An advantage of the RG approach is that it provides clear rules for the required ingredients to reach a given accuracy. At LL, we need the lowest order hard function $\bm{\mathcal{H}}_2$ which in our case involves two hard partons, while the soft function is trivial, i.e. $\bm{\mathcal{S}}_m=\bm{1}$. Hard functions with higher multiplicities contribute at higher order. The running at LL is governed by the one-loop anomalous dimension $\bm{\Gamma}^{(1)}$ which is the leading-order contribution in the perturbative expansion of the anomalous dimension $\bm{\Gamma} = \frac{\alpha_s}{4\pi}\bm{\Gamma}^{(1)}+\left(\frac{\alpha_s}{4\pi}\right)^2\bm{\Gamma}^{(2)}+\mathcal{O}\left(\alpha_s^3\right)$. The one-loop anomalous dimension takes the form
\begin{equation}
\bm{\Gamma}^{(1)}_{mn} =\sum_{[ij]} \left[\bm{V}^{ij}_m \,\delta_{m,n} + \bm{R}^{ij}_m\, \delta_{m,n-1}\right] \,. 
\end{equation}
This anomalous dimension matrix has a simple physical interpretation. The $\bm{V}^{ij}_m$ terms on the diagonal are associated with soft singularities of virtual corrections between legs $i$ and $j$, while the $\bm{R}^{ij}_m$ entries are related to real emissions along the direction $n_q$. In the large-$N_c$ limit, the sum only includes neighboring dipoles, and the color structure of $\bm{\Gamma}^{(1)}$ becomes trivial, which drastically simplifies the solution of \eqref{eq:hrdRG}. In the $\rm\overline{MS}$-scheme we obtain
\begin{align}
\boldsymbol{R}^{ij}_m = &\, + \spac 4\spac N_c \spac\spac \frac{n_{ij}}{n_{iq}\spac n_{jq}}\spac \theta_{\rm in}(n_q) \equiv 4\spac N_c \spac W_{ij}^q\spac\theta_{\rm in}(n_q)\,,\\
\boldsymbol{V}^{ij}_m = & \,- \spac4\spac N_c  \int\! \frac{\mathrm{d}\Omega_q}{4\pi} \spac W_{ij}^q\, \equiv  \,-4\spac N_c  \int [\mathrm{d}^2\Omega_q]  \spac W_{ij}^q\,,
\end{align}
where $n_{ab}=n_a \cdot n_b$ throughout this Letter. The dipole radiator $W_{ij}^q$ is the product of the two eikonal factors. 

To reach NLL accuracy it is necessary to include virtual as well as real corrections to the hard $\bm{\mathcal{H}}_m$ and soft $\bm{\mathcal{S}}_m$ functions along with the two-loop anomalous dimension $\bm{\Gamma}^{(2)}$ which governs the running at subleading accuracy. A result for $\bm{\Gamma}^{(2)}$ was presented in \cite{CaronHuot:2015bja}, but in an unconventional renormalization scheme. The anomalous dimension in the standard $\overline{\rm MS}$ scheme was extracted in \cite{Becher:2021urs} and contains additional terms from expanding angular integrals in $d=4-2\epsilon$ around $\epsilon=0$. In the large-$N_c$ limit it reduces to a sum over dipoles
\begin{equation}
\bm{\Gamma}^{(2)}_{mn} =\sum_{[ij]} \left[\bm{v}^{ij}_m \,\delta_{m,n} + \bm{r}^{ij}_m\, \delta_{m,n-1} + \bm{d}^{\spac ij}_m\, \delta_{m,n-2}\right] \,,
\end{equation}
while the general expression also includes three-leg terms \cite{CaronHuot:2015bja,Becher:2021urs}.
The three entries describe double virtual $\bm{v}^{ij}_m$, real-virtual $\bm{r}^{ij}_m$ and double-real $\bm{d}^{\spac ij}_m$ contributions.
The result for these reads
\begin{equation}
\begin{aligned}
\boldsymbol{d}^{\spac ij}_m = & +N_c \spac\big(K_{ij;qr}+K_{ji;qr}\big) \spac \theta_{\mathrm{in}}(n_q)\spac \spac\theta_{\mathrm{in}}(n_r)\,, \\[2mm]
\boldsymbol{r}^{ij}_m = & -N_c\int \big[d^2\Omega_r\big]\big(K_{ij;qr}+K_{ji;qr}\big)\spac \theta_{\mathrm{in}}(n_q) \\
& \spac +8\spac N_c^2 \int \big[d^2\Omega_r\big] \spac M_{ij;qr} \spac \theta_{\mathrm{in}}(n_q) \\[1mm]
& \spac  + \spac N_c \spac \big( 4 \beta_0 \spac X_{ij}^{q} 
+  \gamma_1^{\mathrm{cusp}}\spac W_{ij}^q \big)  \spac \theta_{\mathrm{in}}(n_q)\,,\\[2mm]
\boldsymbol{v}^{ij}_m  = &  \spac - \spac\spac N_c \int \big[d^2\Omega_q\big] \big( 4 \beta_0 \spac X_{ij}^{q} 
+  \gamma_1^{\mathrm{cusp}}\spac W_{ij}^q \big)\,.\label{eq:gamma2}
\end{aligned}
\end{equation}
The directions of the two real or virtual soft gluons are denoted by $n_q,n_r$. The angular function $K_{ij;qr}$ can be found in \cite{CaronHuot:2015bja,Becher:2021urs}. The remaining functions are $X^{\spac q}_{ij}= W_{ij}^q \spac \ln(4 s_q^2)$ and 
\begin{align}
M_{ij;qr}&= \left(W_{ij}^q W_{ij}^r -W_{ij}^q W_{qj}^r-W_{ij}^r W_{rj}^q\right) \ln\!\frac{s_{qr}^2}{s_q^2}\, , 
\label{eq:stronglyorderedM}
\end{align}
where $s_q$ denotes the sine of the azimuthal angle of $n_q$ in the rest frame of the emitting dipole, and $s_{qr}$ is the sine of the azimuthal angle difference. These functions emerge after taking into account the extra terms derived in \cite{Becher:2021urs}. The resulting angular integrals are Lorentz invariant up to the gap constraints. Manifestly invariant expressions are provided in the Supplemental Material and in \cite{BecherSchalchXuUpcoming}. The invariance allows us to generate the emissions in the back-to-back frame of the emitting dipole and was instrumental in finding efficient parametrizations for sampling the integrals, which is crucial for obtaining reliable MC predictions. We note that the angular functions become singular when additional emissions along $n_q$ and $n_r$ are either collinear to one of the parents $n_i,n_j$ or to each other. To regularize these singularities we impose a cut $\tan(\vartheta/2) > e^{-\eta_{\rm cut}}$ on all angles $\vartheta$ between two directions in the lab frame, with $\eta_{\rm cut}=5$. Once we combine real and virtual contributions the singular regions cancel. We have verified that the remaining cutoff effects are negligible for the values of $Q_0$ we consider. We observe that $M_{ij;qr}$ is collinear finite -- the angular integration over the region where $n_r$ is inside the jet vanishes. 

To perform the resummation, it is necessary to solve the renormalization group equations iteratively to evolve the hard functions from $\mu_h\sim~Q$ to the scale $\mu_s~\sim~Q_0$ associated with soft emissions. We thus calculate
\begin{align}\label{eq:US}
   &\bm{\mathcal{H}}_2(\mu_h)\,\bm{U}_{2m}(\mu_h,\mu_s) 
    = \bm{\mathcal{H}}_2(\mu_h)\,{\bf P}\exp\!\left[ 
    \int_{\mu_s}^{\mu_h}\!\frac{d\mu}{\mu}\,\bm{\Gamma} \right]_{2m} \nonumber\\[2mm]
   &= \bm{\mathcal{H}}_2(t_0) \spac \bm{U}_{2m}(t_0,t) + \bm{\mathcal{H}}_2(t_0) \spac \Delta\bm{U}_{2m}(t_0,t) + \cdots,
\end{align}
where we introduced the evolution time $t\equiv t(\mu_h,\mu_s) = \frac{1}{2\beta_0}\ln\left(\frac{\alpha_s(\mu_s)}{\alpha_s(\mu_h)}\right)$ and denoted the  LL evolution from $k$ to $l$ partons by $\bm{U}_{kl}(t_0,t)$ and the NLL correction through $\Delta\bm{U}_{kl}(t_0,t)$.  We point out that once $\mu_h$ has been fixed, for instance to $\mu_h=M_Z$, the mapping $\mu_s\mapsto t$ is unambiguous. The LL evolution factor
\begin{align}
\bm{U}_{kl}(t_0,t) = {\bf P}\exp\!\left[ (t-t_0)\spac\bm{\Gamma}^{(1)} \right]_{kl}
\end{align}
has been implemented in a large-$N_c$ parton shower framework \cite{Balsiger:2018ezi} following \cite{Dasgupta:2001sh}. To reach next-to-leading logarithmic accuracy we have augmented the LL implementation with exactly one insertion of the two-loop anomalous dimension $\bm{\Gamma}^{(2)}$. This is sufficient to capture the effects of RG running at two loops. We denote this insertion by
\begin{align}
&\Delta\bm{U}_{kl}(t_0,t)  =  \nonumber \\ &\int_{t_0}^t \! dt' \spac\spac\spac \bm{U}_{kk'}(t_0,t')\cdot \frac{\alpha_s(t')}{4\pi}\spac\Big(\bm{\Gamma}^{(2)}_{k'l'}-\frac{\beta_1}{\beta_0} \bm{\Gamma}^{(1)}_{k'l'}\Big)\cdot \bm{U}_{l'l}(t',t).\label{eq:2looprunning}
\end{align}
In short, this means that we start a LL shower $\bm{U}_{kl}(t_0,t')$ which runs from $t_0$ until $t'$, where we evaluate all contributions in \eqref{eq:gamma2}, together with a $\beta_1$-correction to the running of the coupling, and then restart a LL shower. A pictorial representation of our NLL shower can be seen in Figure \ref{fig:shower}. The figure shows an insertion of the two-loop anomalous dimension after the first emission, but in the shower, we insert $\bm{\Gamma}^{(2)}$ after after any number of
 emissions. We provide a detailed description of our MC algorithm to calculate \eqref{eq:2looprunning} in the Supplemental Material. To obtain results at full NLL accuracy, it is necessary to include matching terms as well. More precisely, we need one-loop corrections to $\bm{\mathcal{H}}_2$, the tree-level result for $\bm{\mathcal{H}}_3$ and the one-loop soft functions $\bm{\mathcal{S}}_m$. We expand these perturbatively in $\alpha_s$
 \begin{align}
\bm{\mathcal{H}}_2 & = \sigma_0 \left(\bm{\mathcal{H}}_2^{(0)} + \frac{\alpha_s}{4\pi}\bm{\mathcal{H}}_2^{(1)} + \cdots \right), \label{eq:H2}\\
\bm{\mathcal{H}}_3  &= \sigma_0  \left(\frac{\alpha_s}{4\pi}\bm{\mathcal{H}}_3^{(1)} + \cdots \right),  \label{eq:H3}\\
\bm{\mathcal{S}}_m & = \bm{1} + \frac{\alpha_s}{4\pi} \bm{\mathcal{S}}_m^{(1)} + \cdots \label{eq:Sm1},
\end{align}
where $\sigma_0$ is the leading-order cross section. The hard functions $\bm{\mathcal{H}}_2^{(1)},\bm{\mathcal{H}}_3^{(1)}$ are given by standard QCD amplitudes squared with their infrared singularities subtracted in the $\overline{\rm MS}$ scheme. The one-loop soft functions $\bm{\mathcal{S}}_m^{(1)}$, on the other hand, can be calculated by the shower MC code from the final emission into the gap \cite{Balsiger:2019tne}.\\

\begin{figure}
\includegraphics[scale=0.47]{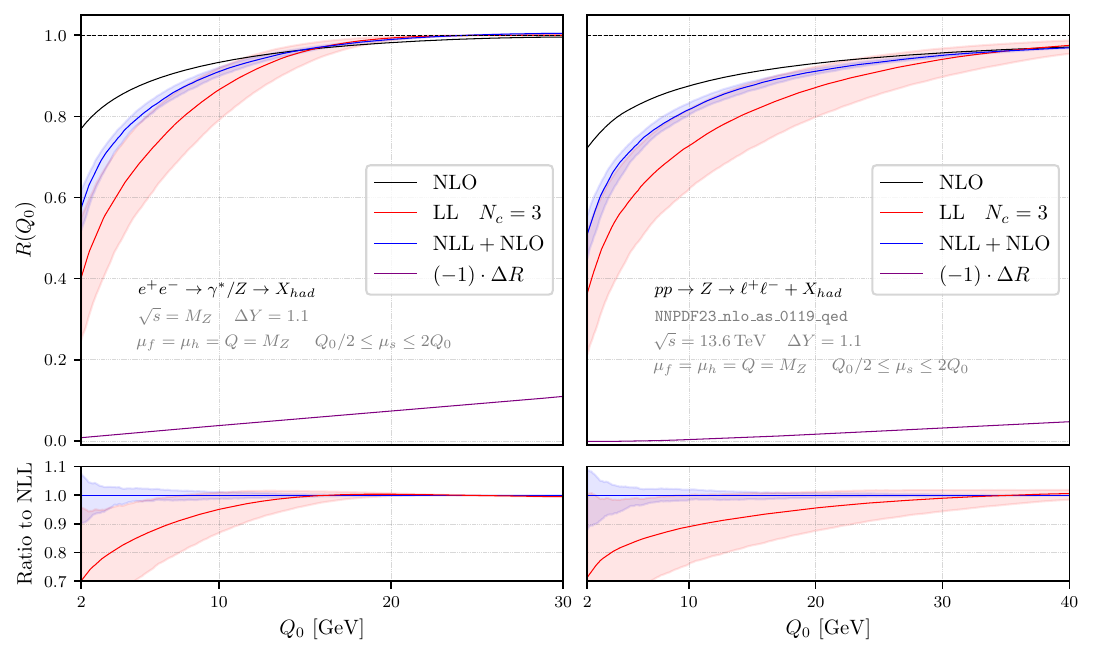}
\caption{\label{fig:NGLs@Mz} Gap-between-jets cross sections at lepton (left) and hadron colliders (right) using $Q=M_Z$. The statistical fluctuations visible in the results are due to the $N_c=3$ LL result.}
\end{figure}
{\em Gap fraction in $e^+e^-$. --- \/}
Analytical expressions for \eqref{eq:H2}-\eqref{eq:Sm1} have been calculated in \cite{Becher:2016mmh} and subsequently implemented in a MC framework \cite{Balsiger:2019tne}. Here we extend results from \cite{Balsiger:2019tne} from $\rm LL'$ to NLL by including corrections to the RG running due to the two-loop anomalous dimension following the aforementioned approach. We work at center-of-mass energy $\sqrt{s}=M_Z$ with $\alpha_s(M_Z)=0.119$ and use two-loop running for $\alpha_s$. Particles are inside the veto region if their angle relative to the thrust axis is larger than $\alpha=\frac{\pi}{3}$, which corresponds to a rapidity gap of $\Delta Y = \ln(3)\approx 1.1$, see Figure~\ref{fig:interjet}. By working with a large angle $\alpha\sim 1$ we avoid collinear logarithms, but the underlying formalism is also available for small angles \cite{Becher:2015hka,Becher:2016mmh}. We calculate the gap fraction 
\begin{align}
R(Q_0)\equiv\frac{1}{\sigma_{\rm tot}}\int_0^{Q_0}dE_T \spac \spac\frac{d\sigma}{dE_T}\,, \label{eq:gapfrac}
\end{align}
which is the fraction of events with transverse energy $E_T$ in the gap below $Q_0$. To include power corrections beyond the factorization theorem \eqref{eq:factthm}, we match to the fixed-order prediction at order $\alpha_s$. The power-suppressed matching corrections $\Delta R$ are included through additive matching. We use profile functions~\cite{Abbate:2010xh} to continuously switch off resummation once the power-suppressed terms become relevant. The shape of the curve at larger $Q_0$ values is affected by this choice and we use the functions introduced in \cite{Balsiger:2019tne}. On the left side of Figure~\ref{fig:NGLs@Mz} we show our numerical results for the gap fraction of the interjet energy flow in $e^+e^-$. The bands arise from varying the soft scale $\mu_s$ since this effect dominates over the $\mu_h$ variation. The LL (red curve) in Figure~\ref{fig:NGLs@Mz} is taken from \cite{Hatta:2013iba} at finite $N_c=3$. After adding the NLL corrections computed from our shower  (blue curve), we obtain a resummed result that is only missing subleading color contributions at NLL. These effects are expected to contribute at the percent level or below. We would like to emphasize two things. First, the large logarithms significantly reduce the gap fraction $R(Q_0)$ in the low-energy regime in comparison to the fixed-order prediction (black line). Secondly, after including the NLL corrections, the width of the scale uncertainty band decreases by a factor of two. We have compared individual ingredients of our computation as well as the full NLL correction to the results of \cite{Banfi:2021owj,Banfi:2021xzn} which are based on a generating functional formalism implemented in the computer code \textsc{Gnole}. We find agreement within numerical uncertainties, as detailed in the Supplemental Material.


\begin{figure}
\includegraphics[scale=0.47]{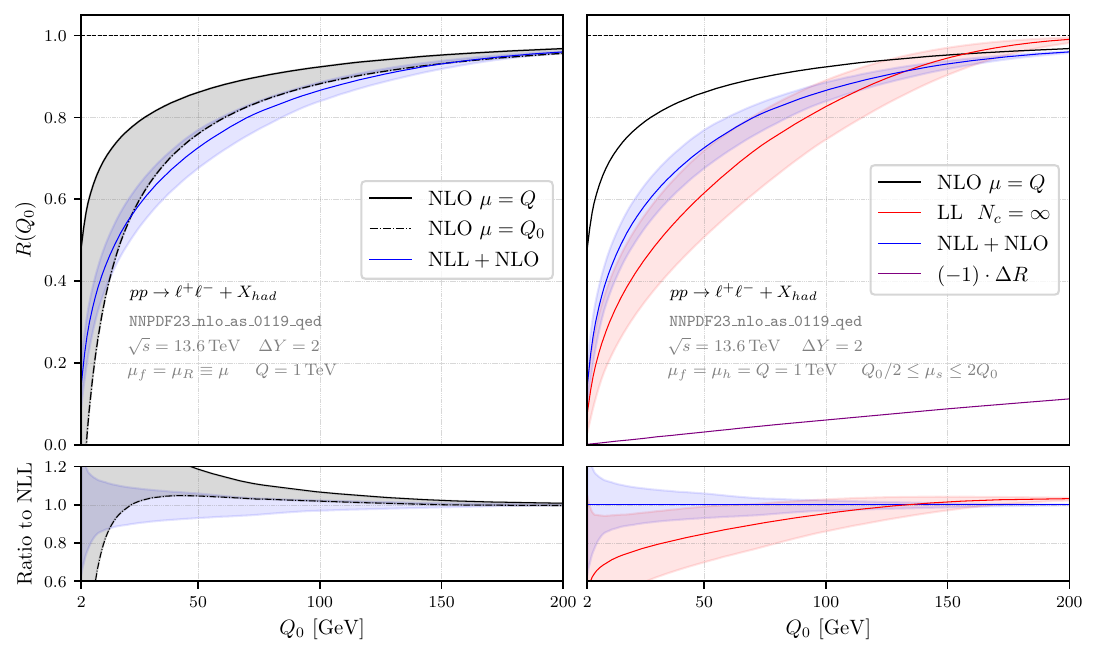}
\caption{\label{fig:NGLs@1TeV} Fixed-order predictions for different scale choices (left) and resummed results at LL and NLL (right) for the value $Q=1\,{\rm TeV}$.}
\end{figure}
{\em Gap fraction in $Z$ production. --- \/} As a first hadron-collider application, we consider $pp \to Z \to \ell^+ \ell^-$ \cite{Drell:1970wh} and compute the cross section for a gap around the incoming beam, centered at the rapidity of the electroweak boson, as considered in \cite{vanBeekveld:2022ukn}. The hadron-collider analog of factorization formula \eqref{eq:factthm} was given in \cite{Balsiger:2018ezi,Becher:2021zkk}. In the large-$N_c$ limit, most ingredients for the resummation carry over from the $e^+e^-$ case, in particular, the RG-evolution and the one-loop soft function can be directly obtained from our MC code. The only new elements are the hard real emission corrections $\bm{\mathcal{H}}_3^{(1)}$. The hard functions are simply the partonic cross sections with their infrared singularities subtracted in the $\overline{\rm MS}$ scheme which are well known \cite{Altarelli:1979ub,Anastasiou:2003yy,Anastasiou:2003ds}. However, the literature assumes that an infrared safe observable is computed and simplifies the cross section results by integrating soft terms over the full angular region. These simplifications are not applicable to the hard functions because the angular constraint in \eqref{eq:hardfct} restricts hard partons to the jet region, while contributions with partons inside the gap are part of the soft function. To avoid a double counting we need to restore the terms which were dropped in \cite{Altarelli:1979ub,Anastasiou:2003yy,Anastasiou:2003ds}. A detailed discussion of this point and explicit expressions for the hard functions in all partonic channels are provided in the Supplemental Material, which includes references \cite{Becher:2010tm,Becher:2007ty,Gribov:1972rt,Dokshitzer:1977sg,Altarelli:1977zs,Lepage:1977sw,Lepage:2020tgj,Gaunt:2014ska,Zeng:2015iba}.

In the right plot of Figure~\ref{fig:NGLs@Mz} we show our numerical results for the gap fraction \eqref{eq:gapfrac}. We work at $\sqrt{s}=13.6$ TeV and use the $\rm NNPDF23\_ nlo\_ as\_0119\_ qed$ \cite{Buckley:2014ana} set and associated $\alpha_s$. We again use the LL results of \cite{Hatta:2013iba} at full color and match to fixed order. We note that the result of \cite{Hatta:2013iba} does not account for complex phase terms in the anomalous dimension \eqref{eq:hrdRG}. These cancel for $e^+e^-$ but are present for hadron colliders and lead to double logarithms at higher orders \cite{Forshaw:2006fk}. However, for $Z$ production the effect of these so-called super-leading logarithms is strongly suppressed and numerically below the percent level \cite{Becher:2021zkk}. For the same gap size $\Delta Y=\ln(3)\approx 1.1$, we see a similar behavior as in the $e^+e^-$ case, see Figure \ref{fig:NGLs@Mz}. To study the effect of large logarithms on the fixed-order expansion, we increase the center-of-mass energy $Q$ to 1~TeV and the gap size $\Delta Y=2$. Once the hierarchy between $Q$ and $Q_0$ is large, it is unclear how the renormalization and factorization scales $\mu_r$ and $\mu_f$ should be chosen in the fixed-order result. In the left plot of Figure~\ref{fig:NGLs@1TeV} we compare the perturbative predictions for a high value $\mu=\mu_f=\mu_r=1\,{\rm TeV}$ (solid black line) and for the choice $\mu=Q_0$ (dashed black line). The difference in the two predictions is an inherent uncertainty in the fixed-order result. The dynamical scale $\mu=Q_0$  leads to better overall agreement with the resummed result, but also to an unphysical negative cross section at low $Q_0$. In our effective field theory approach, each part is evaluated at its natural scale. On the right side of Figure~\ref{fig:NGLs@1TeV} we show the LL as well as the NLL results, both obtained in the large-$N_c$ limit. In all cases, we observe that at low $Q_0$ the NLL effects are around $20\%$ and are within the large LL uncertainty band. As expected, the scale uncertainty is greatly reduced at NLL.
\\

 {\em Conclusions and outlook. --- \/} By performing resummation at subleading logarithmic accuracy, we have reached an important milestone in the effective field theory based resummation of jet observables. The simple non-global observables computed here are of limited phenomenological interest but can serve as benchmark results for the development of general-purpose parton showers at subleading accuracy~\cite{Dasgupta:2020fwr,Forshaw:2020wrq,Nagy:2020dvz,vanBeekveld:2022zhl,Herren:2022jej}. Our MC framework applies directly to more complicated observables. Phenomenologically interesting examples include $H+\mathrm{jets}$ production with a veto on additional jets, and (the background to) Higgs-boson production from vector-boson fusion with a jet veto for central rapidities. The shower evolution is general; the only additional ingredients to the resummation are the relevant NLO hard functions, which should be extracted from existing fixed-order codes. It will also be very interesting to apply our framework to jets with small radii, for example, to compute the jet mass or photon isolation effects \cite{Becher:2022rhu} at subleading logarithmic accuracy. The factorization theorems for these observables are~available~\cite{Becher:2016omr,Becher:2022rhu}.\\

{\em Note added. --- \/} After completion of this work, the Letter \cite{FerrarioRavasio:2023kyg} appeared, which includes the full fermionic contribution in the limit where one counts $n_F\sim N_c$. We have adapted our resummation to capture this contribution as well, see the Supplemental Material for details.\\

 {\em Acknowledgements. --- \/} We thank J\"urg Haag, Dominik Schwienbacher, and Michel Stillger for their comments on the manuscript. We are grateful to Pier Monni for engaging in a detailed comparison, which led to the identification of a problem in an earlier version of our MC code.  We thank Yoshitaka Hatta for sharing his $N_c=3$ results \cite{Hatta:2013iba}. This work was supported by the Swiss National Science Foundation (SNF) under grant 200020\_182038 and by the Cluster of Excellence PRISMA+ (Preci\-
sion Physics, Fundamental Interactions, and Structure of Matter, EXC 2118/1) funded by the German Research Foundation (DFG) under Germany’s Excellence Strategy (Project
ID 390831469). T.B.\ would like to thank the Pauli Center at ETHZ for  hospitality. 
T.B.\ and N.S.\ acknowledge the hospitality of CERN during different stages of this work.  

\bibliographystyle{apsrev4-2}
\bibliography{letterbib}

\begin{thebibliography}{63}%
\makeatletter
\providecommand \@ifxundefined [1]{%
 \@ifx{#1\undefined}
}%
\providecommand \@ifnum [1]{%
 \ifnum #1\expandafter \@firstoftwo
 \else \expandafter \@secondoftwo
 \fi
}%
\providecommand \@ifx [1]{%
 \ifx #1\expandafter \@firstoftwo
 \else \expandafter \@secondoftwo
 \fi
}%
\providecommand \natexlab [1]{#1}%
\providecommand \enquote  [1]{``#1''}%
\providecommand \bibnamefont  [1]{#1}%
\providecommand \bibfnamefont [1]{#1}%
\providecommand \citenamefont [1]{#1}%
\providecommand \href@noop [0]{\@secondoftwo}%
\providecommand \href [0]{\begingroup \@sanitize@url \@href}%
\providecommand \@href[1]{\@@startlink{#1}\@@href}%
\providecommand \@@href[1]{\endgroup#1\@@endlink}%
\providecommand \@sanitize@url [0]{\catcode `\\12\catcode `\$12\catcode
  `\&12\catcode `\#12\catcode `\^12\catcode `\_12\catcode `\%12\relax}%
\providecommand \@@startlink[1]{}%
\providecommand \@@endlink[0]{}%
\providecommand \url  [0]{\begingroup\@sanitize@url \@url }%
\providecommand \@url [1]{\endgroup\@href {#1}{\urlprefix }}%
\providecommand \urlprefix  [0]{URL }%
\providecommand \Eprint [0]{\href }%
\providecommand \doibase [0]{https://doi.org/}%
\providecommand \selectlanguage [0]{\@gobble}%
\providecommand \bibinfo  [0]{\@secondoftwo}%
\providecommand \bibfield  [0]{\@secondoftwo}%
\providecommand \translation [1]{[#1]}%
\providecommand \BibitemOpen [0]{}%
\providecommand \bibitemStop [0]{}%
\providecommand \bibitemNoStop [0]{.\EOS\space}%
\providecommand \EOS [0]{\spacefactor3000\relax}%
\providecommand \BibitemShut  [1]{\csname bibitem#1\endcsname}%
\let\auto@bib@innerbib\@empty
\bibitem [{\citenamefont {Dasgupta}\ and\ \citenamefont
  {Salam}(2001)}]{Dasgupta:2001sh}%
  \BibitemOpen
  \bibfield  {author} {\bibinfo {author} {\bibfnamefont {M.}~\bibnamefont
  {Dasgupta}}\ and\ \bibinfo {author} {\bibfnamefont {G.~P.}\ \bibnamefont
  {Salam}},\ }\href {https://doi.org/10.1016/S0370-2693(01)00725-0} {\bibfield
  {journal} {\bibinfo  {journal} {Phys. Lett. B}\ }\textbf {\bibinfo {volume}
  {512}},\ \bibinfo {pages} {323} (\bibinfo {year} {2001})},\ \Eprint
  {https://arxiv.org/abs/hep-ph/0104277} {arXiv:hep-ph/0104277} \BibitemShut
  {NoStop}%
\bibitem [{\citenamefont {Dasgupta}\ and\ \citenamefont
  {Salam}(2002)}]{Dasgupta:2002bw}%
  \BibitemOpen
  \bibfield  {author} {\bibinfo {author} {\bibfnamefont {M.}~\bibnamefont
  {Dasgupta}}\ and\ \bibinfo {author} {\bibfnamefont {G.~P.}\ \bibnamefont
  {Salam}},\ }\href {https://doi.org/10.1088/1126-6708/2002/03/017} {\bibfield
  {journal} {\bibinfo  {journal} {JHEP}\ }\textbf {\bibinfo {volume} {03}},\
  \bibinfo {pages} {017}},\ \Eprint {https://arxiv.org/abs/hep-ph/0203009}
  {arXiv:hep-ph/0203009} \BibitemShut {NoStop}%
\bibitem [{\citenamefont {Banfi}\ \emph {et~al.}(2002)\citenamefont {Banfi},
  \citenamefont {Marchesini},\ and\ \citenamefont {Smye}}]{Banfi:2002hw}%
  \BibitemOpen
  \bibfield  {author} {\bibinfo {author} {\bibfnamefont {A.}~\bibnamefont
  {Banfi}}, \bibinfo {author} {\bibfnamefont {G.}~\bibnamefont {Marchesini}},\
  and\ \bibinfo {author} {\bibfnamefont {G.}~\bibnamefont {Smye}},\ }\href
  {https://doi.org/10.1088/1126-6708/2002/08/006} {\bibfield  {journal}
  {\bibinfo  {journal} {JHEP}\ }\textbf {\bibinfo {volume} {08}},\ \bibinfo
  {pages} {006}},\ \Eprint {https://arxiv.org/abs/hep-ph/0206076}
  {arXiv:hep-ph/0206076} \BibitemShut {NoStop}%
\bibitem [{\citenamefont {Hatta}\ and\ \citenamefont
  {Ueda}(2013)}]{Hatta:2013iba}%
  \BibitemOpen
  \bibfield  {author} {\bibinfo {author} {\bibfnamefont {Y.}~\bibnamefont
  {Hatta}}\ and\ \bibinfo {author} {\bibfnamefont {T.}~\bibnamefont {Ueda}},\
  }\href {https://doi.org/10.1016/j.nuclphysb.2013.06.021} {\bibfield
  {journal} {\bibinfo  {journal} {Nucl. Phys. B}\ }\textbf {\bibinfo {volume}
  {874}},\ \bibinfo {pages} {808} (\bibinfo {year} {2013})},\ \Eprint
  {https://arxiv.org/abs/1304.6930} {arXiv:1304.6930 [hep-ph]} \BibitemShut
  {NoStop}%
\bibitem [{\citenamefont {Hagiwara}\ \emph {et~al.}(2016)\citenamefont
  {Hagiwara}, \citenamefont {Hatta},\ and\ \citenamefont
  {Ueda}}]{Hagiwara:2015bia}%
  \BibitemOpen
  \bibfield  {author} {\bibinfo {author} {\bibfnamefont {Y.}~\bibnamefont
  {Hagiwara}}, \bibinfo {author} {\bibfnamefont {Y.}~\bibnamefont {Hatta}},\
  and\ \bibinfo {author} {\bibfnamefont {T.}~\bibnamefont {Ueda}},\ }\href
  {https://doi.org/10.1016/j.physletb.2016.03.028} {\bibfield  {journal}
  {\bibinfo  {journal} {Phys. Lett. B}\ }\textbf {\bibinfo {volume} {756}},\
  \bibinfo {pages} {254} (\bibinfo {year} {2016})},\ \Eprint
  {https://arxiv.org/abs/1507.07641} {arXiv:1507.07641 [hep-ph]} \BibitemShut
  {NoStop}%
\bibitem [{\citenamefont {Hatta}\ and\ \citenamefont
  {Ueda}(2021)}]{Hatta:2020wre}%
  \BibitemOpen
  \bibfield  {author} {\bibinfo {author} {\bibfnamefont {Y.}~\bibnamefont
  {Hatta}}\ and\ \bibinfo {author} {\bibfnamefont {T.}~\bibnamefont {Ueda}},\
  }\href {https://doi.org/10.1016/j.nuclphysb.2020.115273} {\bibfield
  {journal} {\bibinfo  {journal} {Nucl. Phys. B}\ }\textbf {\bibinfo {volume}
  {962}},\ \bibinfo {pages} {115273} (\bibinfo {year} {2021})},\ \Eprint
  {https://arxiv.org/abs/2011.04154} {arXiv:2011.04154 [hep-ph]} \BibitemShut
  {NoStop}%
\bibitem [{\citenamefont {De~Angelis}\ \emph {et~al.}(2021)\citenamefont
  {De~Angelis}, \citenamefont {Forshaw},\ and\ \citenamefont
  {Pl\"atzer}}]{DeAngelis:2020rvq}%
  \BibitemOpen
  \bibfield  {author} {\bibinfo {author} {\bibfnamefont {M.}~\bibnamefont
  {De~Angelis}}, \bibinfo {author} {\bibfnamefont {J.~R.}\ \bibnamefont
  {Forshaw}},\ and\ \bibinfo {author} {\bibfnamefont {S.}~\bibnamefont
  {Pl\"atzer}},\ }\href {https://doi.org/10.1103/PhysRevLett.126.112001}
  {\bibfield  {journal} {\bibinfo  {journal} {Phys. Rev. Lett.}\ }\textbf
  {\bibinfo {volume} {126}},\ \bibinfo {pages} {112001} (\bibinfo {year}
  {2021})},\ \Eprint {https://arxiv.org/abs/2007.09648} {arXiv:2007.09648
  [hep-ph]} \BibitemShut {NoStop}%
\bibitem [{\citenamefont {Weigert}(2004)}]{Weigert:2003mm}%
  \BibitemOpen
  \bibfield  {author} {\bibinfo {author} {\bibfnamefont {H.}~\bibnamefont
  {Weigert}},\ }\href {https://doi.org/10.1016/j.nuclphysb.2004.03.002}
  {\bibfield  {journal} {\bibinfo  {journal} {Nucl. Phys. B}\ }\textbf
  {\bibinfo {volume} {685}},\ \bibinfo {pages} {321} (\bibinfo {year}
  {2004})},\ \Eprint {https://arxiv.org/abs/hep-ph/0312050}
  {arXiv:hep-ph/0312050} \BibitemShut {NoStop}%
\bibitem [{\citenamefont {Forshaw}\ \emph {et~al.}(2006)\citenamefont
  {Forshaw}, \citenamefont {Kyrieleis},\ and\ \citenamefont
  {Seymour}}]{Forshaw:2006fk}%
  \BibitemOpen
  \bibfield  {author} {\bibinfo {author} {\bibfnamefont {J.~R.}\ \bibnamefont
  {Forshaw}}, \bibinfo {author} {\bibfnamefont {A.}~\bibnamefont {Kyrieleis}},\
  and\ \bibinfo {author} {\bibfnamefont {M.~H.}\ \bibnamefont {Seymour}},\
  }\href {https://doi.org/10.1088/1126-6708/2006/08/059} {\bibfield  {journal}
  {\bibinfo  {journal} {JHEP}\ }\textbf {\bibinfo {volume} {08}},\ \bibinfo
  {pages} {059}},\ \Eprint {https://arxiv.org/abs/hep-ph/0604094}
  {arXiv:hep-ph/0604094} \BibitemShut {NoStop}%
\bibitem [{\citenamefont {Forshaw}\ \emph {et~al.}(2009)\citenamefont
  {Forshaw}, \citenamefont {Keates},\ and\ \citenamefont
  {Marzani}}]{Forshaw:2009fz}%
  \BibitemOpen
  \bibfield  {author} {\bibinfo {author} {\bibfnamefont {J.}~\bibnamefont
  {Forshaw}}, \bibinfo {author} {\bibfnamefont {J.}~\bibnamefont {Keates}},\
  and\ \bibinfo {author} {\bibfnamefont {S.}~\bibnamefont {Marzani}},\ }\href
  {https://doi.org/10.1088/1126-6708/2009/07/023} {\bibfield  {journal}
  {\bibinfo  {journal} {JHEP}\ }\textbf {\bibinfo {volume} {07}},\ \bibinfo
  {pages} {023}},\ \Eprint {https://arxiv.org/abs/0905.1350} {arXiv:0905.1350
  [hep-ph]} \BibitemShut {NoStop}%
\bibitem [{\citenamefont {Duran~Delgado}\ \emph {et~al.}(2011)\citenamefont
  {Duran~Delgado}, \citenamefont {Forshaw}, \citenamefont {Marzani},\ and\
  \citenamefont {Seymour}}]{DuranDelgado:2011tp}%
  \BibitemOpen
  \bibfield  {author} {\bibinfo {author} {\bibfnamefont {R.~M.}\ \bibnamefont
  {Duran~Delgado}}, \bibinfo {author} {\bibfnamefont {J.~R.}\ \bibnamefont
  {Forshaw}}, \bibinfo {author} {\bibfnamefont {S.}~\bibnamefont {Marzani}},\
  and\ \bibinfo {author} {\bibfnamefont {M.~H.}\ \bibnamefont {Seymour}},\
  }\href {https://doi.org/10.1007/JHEP08(2011)157} {\bibfield  {journal}
  {\bibinfo  {journal} {JHEP}\ }\textbf {\bibinfo {volume} {08}},\ \bibinfo
  {pages} {157}},\ \Eprint {https://arxiv.org/abs/1107.2084} {arXiv:1107.2084
  [hep-ph]} \BibitemShut {NoStop}%
\bibitem [{\citenamefont {Schwartz}\ and\ \citenamefont
  {Zhu}(2014)}]{Schwartz:2014wha}%
  \BibitemOpen
  \bibfield  {author} {\bibinfo {author} {\bibfnamefont {M.~D.}\ \bibnamefont
  {Schwartz}}\ and\ \bibinfo {author} {\bibfnamefont {H.~X.}\ \bibnamefont
  {Zhu}},\ }\href {https://doi.org/10.1103/PhysRevD.90.065004} {\bibfield
  {journal} {\bibinfo  {journal} {Phys. Rev. D}\ }\textbf {\bibinfo {volume}
  {90}},\ \bibinfo {pages} {065004} (\bibinfo {year} {2014})},\ \Eprint
  {https://arxiv.org/abs/1403.4949} {arXiv:1403.4949 [hep-ph]} \BibitemShut
  {NoStop}%
\bibitem [{\citenamefont {Becher}\ \emph
  {et~al.}(2016{\natexlab{a}})\citenamefont {Becher}, \citenamefont {Neubert},
  \citenamefont {Rothen},\ and\ \citenamefont {Shao}}]{Becher:2015hka}%
  \BibitemOpen
  \bibfield  {author} {\bibinfo {author} {\bibfnamefont {T.}~\bibnamefont
  {Becher}}, \bibinfo {author} {\bibfnamefont {M.}~\bibnamefont {Neubert}},
  \bibinfo {author} {\bibfnamefont {L.}~\bibnamefont {Rothen}},\ and\ \bibinfo
  {author} {\bibfnamefont {D.~Y.}\ \bibnamefont {Shao}},\ }\href
  {https://doi.org/10.1103/PhysRevLett.116.192001} {\bibfield  {journal}
  {\bibinfo  {journal} {Phys. Rev. Lett.}\ }\textbf {\bibinfo {volume} {116}},\
  \bibinfo {pages} {192001} (\bibinfo {year} {2016}{\natexlab{a}})},\ \Eprint
  {https://arxiv.org/abs/1508.06645} {arXiv:1508.06645 [hep-ph]} \BibitemShut
  {NoStop}%
\bibitem [{\citenamefont {Becher}\ \emph
  {et~al.}(2016{\natexlab{b}})\citenamefont {Becher}, \citenamefont {Neubert},
  \citenamefont {Rothen},\ and\ \citenamefont {Shao}}]{Becher:2016mmh}%
  \BibitemOpen
  \bibfield  {author} {\bibinfo {author} {\bibfnamefont {T.}~\bibnamefont
  {Becher}}, \bibinfo {author} {\bibfnamefont {M.}~\bibnamefont {Neubert}},
  \bibinfo {author} {\bibfnamefont {L.}~\bibnamefont {Rothen}},\ and\ \bibinfo
  {author} {\bibfnamefont {D.~Y.}\ \bibnamefont {Shao}},\ }\href
  {https://doi.org/10.1007/JHEP11(2016)019} {\bibfield  {journal} {\bibinfo
  {journal} {JHEP}\ }\textbf {\bibinfo {volume} {11}},\ \bibinfo {pages}
  {019}},\ \bibinfo {note} {[Erratum: JHEP 05, 154 (2017)]},\ \Eprint
  {https://arxiv.org/abs/1605.02737} {arXiv:1605.02737 [hep-ph]} \BibitemShut
  {NoStop}%
\bibitem [{\citenamefont {Larkoski}\ \emph {et~al.}(2015)\citenamefont
  {Larkoski}, \citenamefont {Moult},\ and\ \citenamefont
  {Neill}}]{Larkoski:2015zka}%
  \BibitemOpen
  \bibfield  {author} {\bibinfo {author} {\bibfnamefont {A.~J.}\ \bibnamefont
  {Larkoski}}, \bibinfo {author} {\bibfnamefont {I.}~\bibnamefont {Moult}},\
  and\ \bibinfo {author} {\bibfnamefont {D.}~\bibnamefont {Neill}},\ }\href
  {https://doi.org/10.1007/JHEP09(2015)143} {\bibfield  {journal} {\bibinfo
  {journal} {JHEP}\ }\textbf {\bibinfo {volume} {09}},\ \bibinfo {pages}
  {143}},\ \Eprint {https://arxiv.org/abs/1501.04596} {arXiv:1501.04596
  [hep-ph]} \BibitemShut {NoStop}%
\bibitem [{\citenamefont {Caron-Huot}(2018)}]{CaronHuot:2015bja}%
  \BibitemOpen
  \bibfield  {author} {\bibinfo {author} {\bibfnamefont {S.}~\bibnamefont
  {Caron-Huot}},\ }\href {https://doi.org/10.1007/JHEP03(2018)036} {\bibfield
  {journal} {\bibinfo  {journal} {JHEP}\ }\textbf {\bibinfo {volume} {03}},\
  \bibinfo {pages} {036}},\ \Eprint {https://arxiv.org/abs/1501.03754}
  {arXiv:1501.03754 [hep-ph]} \BibitemShut {NoStop}%
\bibitem [{\citenamefont {Becher}\ \emph
  {et~al.}(2016{\natexlab{c}})\citenamefont {Becher}, \citenamefont {Pecjak},\
  and\ \citenamefont {Shao}}]{Becher:2016omr}%
  \BibitemOpen
  \bibfield  {author} {\bibinfo {author} {\bibfnamefont {T.}~\bibnamefont
  {Becher}}, \bibinfo {author} {\bibfnamefont {B.~D.}\ \bibnamefont {Pecjak}},\
  and\ \bibinfo {author} {\bibfnamefont {D.~Y.}\ \bibnamefont {Shao}},\ }\href
  {https://doi.org/10.1007/JHEP12(2016)018} {\bibfield  {journal} {\bibinfo
  {journal} {JHEP}\ }\textbf {\bibinfo {volume} {12}},\ \bibinfo {pages}
  {018}},\ \Eprint {https://arxiv.org/abs/1610.01608} {arXiv:1610.01608
  [hep-ph]} \BibitemShut {NoStop}%
\bibitem [{\citenamefont {Neill}(2017)}]{Neill:2016stq}%
  \BibitemOpen
  \bibfield  {author} {\bibinfo {author} {\bibfnamefont {D.}~\bibnamefont
  {Neill}},\ }\href {https://doi.org/10.1007/JHEP01(2017)109} {\bibfield
  {journal} {\bibinfo  {journal} {JHEP}\ }\textbf {\bibinfo {volume} {01}},\
  \bibinfo {pages} {109}},\ \Eprint {https://arxiv.org/abs/1610.02031}
  {arXiv:1610.02031 [hep-ph]} \BibitemShut {NoStop}%
\bibitem [{\citenamefont {Caron-Huot}\ and\ \citenamefont
  {Herranen}(2018)}]{Caron-Huot:2016tzz}%
  \BibitemOpen
  \bibfield  {author} {\bibinfo {author} {\bibfnamefont {S.}~\bibnamefont
  {Caron-Huot}}\ and\ \bibinfo {author} {\bibfnamefont {M.}~\bibnamefont
  {Herranen}},\ }\href {https://doi.org/10.1007/JHEP02(2018)058} {\bibfield
  {journal} {\bibinfo  {journal} {JHEP}\ }\textbf {\bibinfo {volume} {02}},\
  \bibinfo {pages} {058}},\ \Eprint {https://arxiv.org/abs/1604.07417}
  {arXiv:1604.07417 [hep-ph]} \BibitemShut {NoStop}%
\bibitem [{\citenamefont {Larkoski}\ \emph {et~al.}(2016)\citenamefont
  {Larkoski}, \citenamefont {Moult},\ and\ \citenamefont
  {Neill}}]{Larkoski:2016zzc}%
  \BibitemOpen
  \bibfield  {author} {\bibinfo {author} {\bibfnamefont {A.~J.}\ \bibnamefont
  {Larkoski}}, \bibinfo {author} {\bibfnamefont {I.}~\bibnamefont {Moult}},\
  and\ \bibinfo {author} {\bibfnamefont {D.}~\bibnamefont {Neill}},\ }\href
  {https://doi.org/10.1007/JHEP11(2016)089} {\bibfield  {journal} {\bibinfo
  {journal} {JHEP}\ }\textbf {\bibinfo {volume} {11}},\ \bibinfo {pages}
  {089}},\ \Eprint {https://arxiv.org/abs/1609.04011} {arXiv:1609.04011
  [hep-ph]} \BibitemShut {NoStop}%
\bibitem [{\citenamefont {Hatta}\ \emph {et~al.}(2018)\citenamefont {Hatta},
  \citenamefont {Iancu}, \citenamefont {Mueller},\ and\ \citenamefont
  {Triantafyllopoulos}}]{Hatta:2017fwr}%
  \BibitemOpen
  \bibfield  {author} {\bibinfo {author} {\bibfnamefont {Y.}~\bibnamefont
  {Hatta}}, \bibinfo {author} {\bibfnamefont {E.}~\bibnamefont {Iancu}},
  \bibinfo {author} {\bibfnamefont {A.~H.}\ \bibnamefont {Mueller}},\ and\
  \bibinfo {author} {\bibfnamefont {D.~N.}\ \bibnamefont
  {Triantafyllopoulos}},\ }\href {https://doi.org/10.1007/JHEP02(2018)075}
  {\bibfield  {journal} {\bibinfo  {journal} {JHEP}\ }\textbf {\bibinfo
  {volume} {02}},\ \bibinfo {pages} {075}},\ \Eprint
  {https://arxiv.org/abs/1710.06722} {arXiv:1710.06722 [hep-ph]} \BibitemShut
  {NoStop}%
\bibitem [{\citenamefont {Becher}\ \emph {et~al.}(2017)\citenamefont {Becher},
  \citenamefont {Rahn},\ and\ \citenamefont {Shao}}]{Becher:2017nof}%
  \BibitemOpen
  \bibfield  {author} {\bibinfo {author} {\bibfnamefont {T.}~\bibnamefont
  {Becher}}, \bibinfo {author} {\bibfnamefont {R.}~\bibnamefont {Rahn}},\ and\
  \bibinfo {author} {\bibfnamefont {D.~Y.}\ \bibnamefont {Shao}},\ }\href
  {https://doi.org/10.1007/JHEP10(2017)030} {\bibfield  {journal} {\bibinfo
  {journal} {JHEP}\ }\textbf {\bibinfo {volume} {10}},\ \bibinfo {pages}
  {030}},\ \Eprint {https://arxiv.org/abs/1708.04516} {arXiv:1708.04516
  [hep-ph]} \BibitemShut {NoStop}%
\bibitem [{\citenamefont {\'Angeles~Mart\'\i{}nez}\ \emph
  {et~al.}(2018)\citenamefont {\'Angeles~Mart\'\i{}nez}, \citenamefont
  {De~Angelis}, \citenamefont {Forshaw}, \citenamefont {Pl\"atzer},\ and\
  \citenamefont {Seymour}}]{AngelesMartinez:2018cfz}%
  \BibitemOpen
  \bibfield  {author} {\bibinfo {author} {\bibfnamefont {R.}~\bibnamefont
  {\'Angeles~Mart\'\i{}nez}}, \bibinfo {author} {\bibfnamefont
  {M.}~\bibnamefont {De~Angelis}}, \bibinfo {author} {\bibfnamefont {J.~R.}\
  \bibnamefont {Forshaw}}, \bibinfo {author} {\bibfnamefont {S.}~\bibnamefont
  {Pl\"atzer}},\ and\ \bibinfo {author} {\bibfnamefont {M.~H.}\ \bibnamefont
  {Seymour}},\ }\href {https://doi.org/10.1007/JHEP05(2018)044} {\bibfield
  {journal} {\bibinfo  {journal} {JHEP}\ }\textbf {\bibinfo {volume} {05}},\
  \bibinfo {pages} {044}},\ \Eprint {https://arxiv.org/abs/1802.08531}
  {arXiv:1802.08531 [hep-ph]} \BibitemShut {NoStop}%
\bibitem [{\citenamefont {Neill}(2019)}]{Neill:2018yet}%
  \BibitemOpen
  \bibfield  {author} {\bibinfo {author} {\bibfnamefont {D.}~\bibnamefont
  {Neill}},\ }\href {https://doi.org/10.1007/JHEP02(2019)114} {\bibfield
  {journal} {\bibinfo  {journal} {JHEP}\ }\textbf {\bibinfo {volume} {02}},\
  \bibinfo {pages} {114}},\ \Eprint {https://arxiv.org/abs/1808.04897}
  {arXiv:1808.04897 [hep-ph]} \BibitemShut {NoStop}%
\bibitem [{\citenamefont {Balsiger}\ \emph {et~al.}(2018)\citenamefont
  {Balsiger}, \citenamefont {Becher},\ and\ \citenamefont
  {Shao}}]{Balsiger:2018ezi}%
  \BibitemOpen
  \bibfield  {author} {\bibinfo {author} {\bibfnamefont {M.}~\bibnamefont
  {Balsiger}}, \bibinfo {author} {\bibfnamefont {T.}~\bibnamefont {Becher}},\
  and\ \bibinfo {author} {\bibfnamefont {D.~Y.}\ \bibnamefont {Shao}},\ }\href
  {https://doi.org/10.1007/JHEP08(2018)104} {\bibfield  {journal} {\bibinfo
  {journal} {JHEP}\ }\textbf {\bibinfo {volume} {08}},\ \bibinfo {pages}
  {104}},\ \Eprint {https://arxiv.org/abs/1803.07045} {arXiv:1803.07045
  [hep-ph]} \BibitemShut {NoStop}%
\bibitem [{\citenamefont {Dreyer}\ \emph {et~al.}(2018)\citenamefont {Dreyer},
  \citenamefont {Salam},\ and\ \citenamefont {Soyez}}]{Dreyer:2018nbf}%
  \BibitemOpen
  \bibfield  {author} {\bibinfo {author} {\bibfnamefont {F.~A.}\ \bibnamefont
  {Dreyer}}, \bibinfo {author} {\bibfnamefont {G.~P.}\ \bibnamefont {Salam}},\
  and\ \bibinfo {author} {\bibfnamefont {G.}~\bibnamefont {Soyez}},\ }\href
  {https://doi.org/10.1007/JHEP12(2018)064} {\bibfield  {journal} {\bibinfo
  {journal} {JHEP}\ }\textbf {\bibinfo {volume} {12}},\ \bibinfo {pages}
  {064}},\ \Eprint {https://arxiv.org/abs/1807.04758} {arXiv:1807.04758
  [hep-ph]} \BibitemShut {NoStop}%
\bibitem [{\citenamefont {Balsiger}\ \emph {et~al.}(2019)\citenamefont
  {Balsiger}, \citenamefont {Becher},\ and\ \citenamefont
  {Shao}}]{Balsiger:2019tne}%
  \BibitemOpen
  \bibfield  {author} {\bibinfo {author} {\bibfnamefont {M.}~\bibnamefont
  {Balsiger}}, \bibinfo {author} {\bibfnamefont {T.}~\bibnamefont {Becher}},\
  and\ \bibinfo {author} {\bibfnamefont {D.~Y.}\ \bibnamefont {Shao}},\ }\href
  {https://doi.org/10.1007/JHEP04(2019)020} {\bibfield  {journal} {\bibinfo
  {journal} {JHEP}\ }\textbf {\bibinfo {volume} {04}},\ \bibinfo {pages}
  {020}},\ \Eprint {https://arxiv.org/abs/1901.09038} {arXiv:1901.09038
  [hep-ph]} \BibitemShut {NoStop}%
\bibitem [{\citenamefont {Balsiger}\ \emph {et~al.}(2020)\citenamefont
  {Balsiger}, \citenamefont {Becher},\ and\ \citenamefont
  {Ferroglia}}]{Balsiger:2020ogy}%
  \BibitemOpen
  \bibfield  {author} {\bibinfo {author} {\bibfnamefont {M.}~\bibnamefont
  {Balsiger}}, \bibinfo {author} {\bibfnamefont {T.}~\bibnamefont {Becher}},\
  and\ \bibinfo {author} {\bibfnamefont {A.}~\bibnamefont {Ferroglia}},\ }\href
  {https://doi.org/10.1007/JHEP09(2020)029} {\bibfield  {journal} {\bibinfo
  {journal} {JHEP}\ }\textbf {\bibinfo {volume} {09}},\ \bibinfo {pages}
  {029}},\ \Eprint {https://arxiv.org/abs/2006.00014} {arXiv:2006.00014
  [hep-ph]} \BibitemShut {NoStop}%
\bibitem [{\citenamefont {Dasgupta}\ \emph {et~al.}(2020)\citenamefont
  {Dasgupta}, \citenamefont {Dreyer}, \citenamefont {Hamilton}, \citenamefont
  {Monni}, \citenamefont {Salam},\ and\ \citenamefont
  {Soyez}}]{Dasgupta:2020fwr}%
  \BibitemOpen
  \bibfield  {author} {\bibinfo {author} {\bibfnamefont {M.}~\bibnamefont
  {Dasgupta}}, \bibinfo {author} {\bibfnamefont {F.~A.}\ \bibnamefont
  {Dreyer}}, \bibinfo {author} {\bibfnamefont {K.}~\bibnamefont {Hamilton}},
  \bibinfo {author} {\bibfnamefont {P.~F.}\ \bibnamefont {Monni}}, \bibinfo
  {author} {\bibfnamefont {G.~P.}\ \bibnamefont {Salam}},\ and\ \bibinfo
  {author} {\bibfnamefont {G.}~\bibnamefont {Soyez}},\ }\href
  {https://doi.org/10.1103/PhysRevLett.125.052002} {\bibfield  {journal}
  {\bibinfo  {journal} {Phys. Rev. Lett.}\ }\textbf {\bibinfo {volume} {125}},\
  \bibinfo {pages} {052002} (\bibinfo {year} {2020})},\ \Eprint
  {https://arxiv.org/abs/2002.11114} {arXiv:2002.11114 [hep-ph]} \BibitemShut
  {NoStop}%
\bibitem [{\citenamefont {Hamilton}\ \emph {et~al.}(2021)\citenamefont
  {Hamilton}, \citenamefont {Medves}, \citenamefont {Salam}, \citenamefont
  {Scyboz},\ and\ \citenamefont {Soyez}}]{Hamilton:2020rcu}%
  \BibitemOpen
  \bibfield  {author} {\bibinfo {author} {\bibfnamefont {K.}~\bibnamefont
  {Hamilton}}, \bibinfo {author} {\bibfnamefont {R.}~\bibnamefont {Medves}},
  \bibinfo {author} {\bibfnamefont {G.~P.}\ \bibnamefont {Salam}}, \bibinfo
  {author} {\bibfnamefont {L.}~\bibnamefont {Scyboz}},\ and\ \bibinfo {author}
  {\bibfnamefont {G.}~\bibnamefont {Soyez}},\ }\href
  {https://doi.org/10.1007/JHEP03(2021)041} {\bibfield  {journal} {\bibinfo
  {journal} {JHEP}\ }\textbf {\bibinfo {volume} {03}}\bibfield  {number}
  {\bibinfo  {number} { (041)},\ \bibinfo {pages} {041}},\ }\Eprint
  {https://arxiv.org/abs/2011.10054} {arXiv:2011.10054 [hep-ph]} \BibitemShut
  {NoStop}%
\bibitem [{\citenamefont {Caletti}\ \emph {et~al.}(2021)\citenamefont
  {Caletti}, \citenamefont {Fedkevych}, \citenamefont {Marzani}, \citenamefont
  {Reichelt}, \citenamefont {Schumann}, \citenamefont {Soyez},\ and\
  \citenamefont {Theeuwes}}]{Caletti:2021oor}%
  \BibitemOpen
  \bibfield  {author} {\bibinfo {author} {\bibfnamefont {S.}~\bibnamefont
  {Caletti}}, \bibinfo {author} {\bibfnamefont {O.}~\bibnamefont {Fedkevych}},
  \bibinfo {author} {\bibfnamefont {S.}~\bibnamefont {Marzani}}, \bibinfo
  {author} {\bibfnamefont {D.}~\bibnamefont {Reichelt}}, \bibinfo {author}
  {\bibfnamefont {S.}~\bibnamefont {Schumann}}, \bibinfo {author}
  {\bibfnamefont {G.}~\bibnamefont {Soyez}},\ and\ \bibinfo {author}
  {\bibfnamefont {V.}~\bibnamefont {Theeuwes}},\ }\href
  {https://doi.org/10.1007/JHEP07(2021)076} {\bibfield  {journal} {\bibinfo
  {journal} {JHEP}\ }\textbf {\bibinfo {volume} {07}},\ \bibinfo {pages}
  {076}},\ \Eprint {https://arxiv.org/abs/2104.06920} {arXiv:2104.06920
  [hep-ph]} \BibitemShut {NoStop}%
\bibitem [{\citenamefont {Becher}\ \emph {et~al.}(2021)\citenamefont {Becher},
  \citenamefont {Neubert},\ and\ \citenamefont {Shao}}]{Becher:2021zkk}%
  \BibitemOpen
  \bibfield  {author} {\bibinfo {author} {\bibfnamefont {T.}~\bibnamefont
  {Becher}}, \bibinfo {author} {\bibfnamefont {M.}~\bibnamefont {Neubert}},\
  and\ \bibinfo {author} {\bibfnamefont {D.~Y.}\ \bibnamefont {Shao}},\ }\href
  {https://doi.org/10.1103/PhysRevLett.127.212002} {\bibfield  {journal}
  {\bibinfo  {journal} {Phys. Rev. Lett.}\ }\textbf {\bibinfo {volume} {127}},\
  \bibinfo {pages} {212002} (\bibinfo {year} {2021})},\ \Eprint
  {https://arxiv.org/abs/2107.01212} {arXiv:2107.01212 [hep-ph]} \BibitemShut
  {NoStop}%
\bibitem [{\citenamefont {van Beekveld}\ \emph
  {et~al.}(2022{\natexlab{a}})\citenamefont {van Beekveld}, \citenamefont
  {Ferrario~Ravasio}, \citenamefont {Salam}, \citenamefont {Soto-Ontoso},
  \citenamefont {Soyez},\ and\ \citenamefont {Verheyen}}]{vanBeekveld:2022zhl}%
  \BibitemOpen
  \bibfield  {author} {\bibinfo {author} {\bibfnamefont {M.}~\bibnamefont {van
  Beekveld}}, \bibinfo {author} {\bibfnamefont {S.}~\bibnamefont
  {Ferrario~Ravasio}}, \bibinfo {author} {\bibfnamefont {G.~P.}\ \bibnamefont
  {Salam}}, \bibinfo {author} {\bibfnamefont {A.}~\bibnamefont {Soto-Ontoso}},
  \bibinfo {author} {\bibfnamefont {G.}~\bibnamefont {Soyez}},\ and\ \bibinfo
  {author} {\bibfnamefont {R.}~\bibnamefont {Verheyen}},\ }\href
  {https://doi.org/10.1007/JHEP11(2022)019} {\bibfield  {journal} {\bibinfo
  {journal} {JHEP}\ }\textbf {\bibinfo {volume} {11}},\ \bibinfo {pages}
  {019}},\ \Eprint {https://arxiv.org/abs/2205.02237} {arXiv:2205.02237
  [hep-ph]} \BibitemShut {NoStop}%
\bibitem [{\citenamefont {van Beekveld}\ \emph
  {et~al.}(2022{\natexlab{b}})\citenamefont {van Beekveld}, \citenamefont
  {Ferrario~Ravasio}, \citenamefont {Hamilton}, \citenamefont {Salam},
  \citenamefont {Soto-Ontoso}, \citenamefont {Soyez},\ and\ \citenamefont
  {Verheyen}}]{vanBeekveld:2022ukn}%
  \BibitemOpen
  \bibfield  {author} {\bibinfo {author} {\bibfnamefont {M.}~\bibnamefont {van
  Beekveld}}, \bibinfo {author} {\bibfnamefont {S.}~\bibnamefont
  {Ferrario~Ravasio}}, \bibinfo {author} {\bibfnamefont {K.}~\bibnamefont
  {Hamilton}}, \bibinfo {author} {\bibfnamefont {G.~P.}\ \bibnamefont {Salam}},
  \bibinfo {author} {\bibfnamefont {A.}~\bibnamefont {Soto-Ontoso}}, \bibinfo
  {author} {\bibfnamefont {G.}~\bibnamefont {Soyez}},\ and\ \bibinfo {author}
  {\bibfnamefont {R.}~\bibnamefont {Verheyen}},\ }\href
  {https://doi.org/10.1007/JHEP11(2022)020} {\bibfield  {journal} {\bibinfo
  {journal} {JHEP}\ }\textbf {\bibinfo {volume} {11}},\ \bibinfo {pages}
  {020}},\ \Eprint {https://arxiv.org/abs/2207.09467} {arXiv:2207.09467
  [hep-ph]} \BibitemShut {NoStop}%
\bibitem [{\citenamefont {van Beekveld}\ and\ \citenamefont
  {Ravasio}(2023)}]{vanBeekveld:2023lfu}%
  \BibitemOpen
  \bibfield  {author} {\bibinfo {author} {\bibfnamefont {M.}~\bibnamefont {van
  Beekveld}}\ and\ \bibinfo {author} {\bibfnamefont {S.~F.}\ \bibnamefont
  {Ravasio}},\ }\href@noop {} {\  (\bibinfo {year} {2023})},\ \Eprint
  {https://arxiv.org/abs/2305.08645} {arXiv:2305.08645 [hep-ph]} \BibitemShut
  {NoStop}%
\bibitem [{\citenamefont {Bauer}\ \emph
  {et~al.}(2002{\natexlab{a}})\citenamefont {Bauer}, \citenamefont {Pirjol},\
  and\ \citenamefont {Stewart}}]{Bauer:2001yt}%
  \BibitemOpen
  \bibfield  {author} {\bibinfo {author} {\bibfnamefont {C.~W.}\ \bibnamefont
  {Bauer}}, \bibinfo {author} {\bibfnamefont {D.}~\bibnamefont {Pirjol}},\ and\
  \bibinfo {author} {\bibfnamefont {I.~W.}\ \bibnamefont {Stewart}},\ }\href
  {https://doi.org/10.1103/PhysRevD.65.054022} {\bibfield  {journal} {\bibinfo
  {journal} {Phys. Rev. D}\ }\textbf {\bibinfo {volume} {65}},\ \bibinfo
  {pages} {054022} (\bibinfo {year} {2002}{\natexlab{a}})},\ \Eprint
  {https://arxiv.org/abs/hep-ph/0109045} {arXiv:hep-ph/0109045} \BibitemShut
  {NoStop}%
\bibitem [{\citenamefont {Bauer}\ \emph
  {et~al.}(2002{\natexlab{b}})\citenamefont {Bauer}, \citenamefont {Fleming},
  \citenamefont {Pirjol}, \citenamefont {Rothstein},\ and\ \citenamefont
  {Stewart}}]{Bauer:2002nz}%
  \BibitemOpen
  \bibfield  {author} {\bibinfo {author} {\bibfnamefont {C.~W.}\ \bibnamefont
  {Bauer}}, \bibinfo {author} {\bibfnamefont {S.}~\bibnamefont {Fleming}},
  \bibinfo {author} {\bibfnamefont {D.}~\bibnamefont {Pirjol}}, \bibinfo
  {author} {\bibfnamefont {I.~Z.}\ \bibnamefont {Rothstein}},\ and\ \bibinfo
  {author} {\bibfnamefont {I.~W.}\ \bibnamefont {Stewart}},\ }\href
  {https://doi.org/10.1103/PhysRevD.66.014017} {\bibfield  {journal} {\bibinfo
  {journal} {Phys. Rev. D}\ }\textbf {\bibinfo {volume} {66}},\ \bibinfo
  {pages} {014017} (\bibinfo {year} {2002}{\natexlab{b}})},\ \Eprint
  {https://arxiv.org/abs/hep-ph/0202088} {arXiv:hep-ph/0202088} \BibitemShut
  {NoStop}%
\bibitem [{\citenamefont {Beneke}\ \emph {et~al.}(2002)\citenamefont {Beneke},
  \citenamefont {Chapovsky}, \citenamefont {Diehl},\ and\ \citenamefont
  {Feldmann}}]{Beneke:2002ph}%
  \BibitemOpen
  \bibfield  {author} {\bibinfo {author} {\bibfnamefont {M.}~\bibnamefont
  {Beneke}}, \bibinfo {author} {\bibfnamefont {A.~P.}\ \bibnamefont
  {Chapovsky}}, \bibinfo {author} {\bibfnamefont {M.}~\bibnamefont {Diehl}},\
  and\ \bibinfo {author} {\bibfnamefont {T.}~\bibnamefont {Feldmann}},\ }\href
  {https://doi.org/10.1016/S0550-3213(02)00687-9} {\bibfield  {journal}
  {\bibinfo  {journal} {Nucl. Phys. B}\ }\textbf {\bibinfo {volume} {643}},\
  \bibinfo {pages} {431} (\bibinfo {year} {2002})},\ \Eprint
  {https://arxiv.org/abs/hep-ph/0206152} {arXiv:hep-ph/0206152} \BibitemShut
  {NoStop}%
\bibitem [{\citenamefont {Becher}\ \emph {et~al.}(2022)\citenamefont {Becher},
  \citenamefont {Rauh},\ and\ \citenamefont {Xu}}]{Becher:2021urs}%
  \BibitemOpen
  \bibfield  {author} {\bibinfo {author} {\bibfnamefont {T.}~\bibnamefont
  {Becher}}, \bibinfo {author} {\bibfnamefont {T.}~\bibnamefont {Rauh}},\ and\
  \bibinfo {author} {\bibfnamefont {X.}~\bibnamefont {Xu}},\ }\href
  {https://doi.org/10.1007/JHEP08(2022)134} {\bibfield  {journal} {\bibinfo
  {journal} {JHEP}\ }\textbf {\bibinfo {volume} {08}},\ \bibinfo {pages}
  {134}},\ \Eprint {https://arxiv.org/abs/2112.02108} {arXiv:2112.02108
  [hep-ph]} \BibitemShut {NoStop}%
\bibitem [{\citenamefont {Banfi}\ \emph {et~al.}(2022)\citenamefont {Banfi},
  \citenamefont {Dreyer},\ and\ \citenamefont {Monni}}]{Banfi:2021xzn}%
  \BibitemOpen
  \bibfield  {author} {\bibinfo {author} {\bibfnamefont {A.}~\bibnamefont
  {Banfi}}, \bibinfo {author} {\bibfnamefont {F.~A.}\ \bibnamefont {Dreyer}},\
  and\ \bibinfo {author} {\bibfnamefont {P.~F.}\ \bibnamefont {Monni}},\ }\href
  {https://doi.org/10.1007/JHEP03(2022)135} {\bibfield  {journal} {\bibinfo
  {journal} {JHEP}\ }\textbf {\bibinfo {volume} {03}},\ \bibinfo {pages}
  {135}},\ \Eprint {https://arxiv.org/abs/2111.02413} {arXiv:2111.02413
  [hep-ph]} \BibitemShut {NoStop}%
\bibitem [{\citenamefont {Banfi}\ \emph {et~al.}(2021)\citenamefont {Banfi},
  \citenamefont {Dreyer},\ and\ \citenamefont {Monni}}]{Banfi:2021owj}%
  \BibitemOpen
  \bibfield  {author} {\bibinfo {author} {\bibfnamefont {A.}~\bibnamefont
  {Banfi}}, \bibinfo {author} {\bibfnamefont {F.~A.}\ \bibnamefont {Dreyer}},\
  and\ \bibinfo {author} {\bibfnamefont {P.~F.}\ \bibnamefont {Monni}},\ }\href
  {https://doi.org/10.1007/JHEP10(2021)006} {\bibfield  {journal} {\bibinfo
  {journal} {JHEP}\ }\textbf {\bibinfo {volume} {10}},\ \bibinfo {pages}
  {006}},\ \Eprint {https://arxiv.org/abs/2104.06416} {arXiv:2104.06416
  [hep-ph]} \BibitemShut {NoStop}%
\bibitem [{\citenamefont {Catani}\ and\ \citenamefont
  {Seymour}(1997)}]{Catani:1996vz}%
  \BibitemOpen
  \bibfield  {author} {\bibinfo {author} {\bibfnamefont {S.}~\bibnamefont
  {Catani}}\ and\ \bibinfo {author} {\bibfnamefont {M.~H.}\ \bibnamefont
  {Seymour}},\ }\href {https://doi.org/10.1016/S0550-3213(96)00589-5}
  {\bibfield  {journal} {\bibinfo  {journal} {Nucl. Phys. B}\ }\textbf
  {\bibinfo {volume} {485}},\ \bibinfo {pages} {291} (\bibinfo {year}
  {1997})},\ \bibinfo {note} {[Erratum: Nucl.Phys.B 510, 503--504 (1998)]},\
  \Eprint {https://arxiv.org/abs/hep-ph/9605323} {arXiv:hep-ph/9605323}
  \BibitemShut {NoStop}%
\bibitem [{\citenamefont {Becher}\ \emph {et~al.}()\citenamefont {Becher},
  \citenamefont {Schalch},\ and\ \citenamefont {Xu}}]{BecherSchalchXuUpcoming}%
  \BibitemOpen
  \bibfield  {author} {\bibinfo {author} {\bibfnamefont {T.}~\bibnamefont
  {Becher}}, \bibinfo {author} {\bibfnamefont {N.}~\bibnamefont {Schalch}},\
  and\ \bibinfo {author} {\bibfnamefont {X.}~\bibnamefont {Xu}},\ }\bibinfo
  {note} {in preparation}\BibitemShut {NoStop}%
\bibitem [{\citenamefont {Abbate}\ \emph {et~al.}(2011)\citenamefont {Abbate},
  \citenamefont {Fickinger}, \citenamefont {Hoang}, \citenamefont {Mateu},\
  and\ \citenamefont {Stewart}}]{Abbate:2010xh}%
  \BibitemOpen
  \bibfield  {author} {\bibinfo {author} {\bibfnamefont {R.}~\bibnamefont
  {Abbate}}, \bibinfo {author} {\bibfnamefont {M.}~\bibnamefont {Fickinger}},
  \bibinfo {author} {\bibfnamefont {A.~H.}\ \bibnamefont {Hoang}}, \bibinfo
  {author} {\bibfnamefont {V.}~\bibnamefont {Mateu}},\ and\ \bibinfo {author}
  {\bibfnamefont {I.~W.}\ \bibnamefont {Stewart}},\ }\href
  {https://doi.org/10.1103/PhysRevD.83.074021} {\bibfield  {journal} {\bibinfo
  {journal} {Phys. Rev. D}\ }\textbf {\bibinfo {volume} {83}},\ \bibinfo
  {pages} {074021} (\bibinfo {year} {2011})},\ \Eprint
  {https://arxiv.org/abs/1006.3080} {arXiv:1006.3080 [hep-ph]} \BibitemShut
  {NoStop}%
\bibitem [{\citenamefont {Drell}\ and\ \citenamefont
  {Yan}(1970)}]{Drell:1970wh}%
  \BibitemOpen
  \bibfield  {author} {\bibinfo {author} {\bibfnamefont {S.~D.}\ \bibnamefont
  {Drell}}\ and\ \bibinfo {author} {\bibfnamefont {T.-M.}\ \bibnamefont
  {Yan}},\ }\href {https://doi.org/10.1103/PhysRevLett.25.316} {\bibfield
  {journal} {\bibinfo  {journal} {Phys. Rev. Lett.}\ }\textbf {\bibinfo
  {volume} {25}},\ \bibinfo {pages} {316} (\bibinfo {year} {1970})},\ \bibinfo
  {note} {[Erratum: Phys.Rev.Lett. 25, 902 (1970)]}\BibitemShut {NoStop}%
\bibitem [{\citenamefont {Altarelli}\ \emph {et~al.}(1979)\citenamefont
  {Altarelli}, \citenamefont {Ellis},\ and\ \citenamefont
  {Martinelli}}]{Altarelli:1979ub}%
  \BibitemOpen
  \bibfield  {author} {\bibinfo {author} {\bibfnamefont {G.}~\bibnamefont
  {Altarelli}}, \bibinfo {author} {\bibfnamefont {R.~K.}\ \bibnamefont
  {Ellis}},\ and\ \bibinfo {author} {\bibfnamefont {G.}~\bibnamefont
  {Martinelli}},\ }\href {https://doi.org/10.1016/0550-3213(79)90116-0}
  {\bibfield  {journal} {\bibinfo  {journal} {Nucl. Phys. B}\ }\textbf
  {\bibinfo {volume} {157}},\ \bibinfo {pages} {461} (\bibinfo {year}
  {1979})}\BibitemShut {NoStop}%
\bibitem [{\citenamefont {Anastasiou}\ \emph {et~al.}(2003)\citenamefont
  {Anastasiou}, \citenamefont {Dixon}, \citenamefont {Melnikov},\ and\
  \citenamefont {Petriello}}]{Anastasiou:2003yy}%
  \BibitemOpen
  \bibfield  {author} {\bibinfo {author} {\bibfnamefont {C.}~\bibnamefont
  {Anastasiou}}, \bibinfo {author} {\bibfnamefont {L.~J.}\ \bibnamefont
  {Dixon}}, \bibinfo {author} {\bibfnamefont {K.}~\bibnamefont {Melnikov}},\
  and\ \bibinfo {author} {\bibfnamefont {F.}~\bibnamefont {Petriello}},\ }\href
  {https://doi.org/10.1103/PhysRevLett.91.182002} {\bibfield  {journal}
  {\bibinfo  {journal} {Phys. Rev. Lett.}\ }\textbf {\bibinfo {volume} {91}},\
  \bibinfo {pages} {182002} (\bibinfo {year} {2003})},\ \Eprint
  {https://arxiv.org/abs/hep-ph/0306192} {arXiv:hep-ph/0306192} \BibitemShut
  {NoStop}%
\bibitem [{\citenamefont {Anastasiou}\ \emph {et~al.}(2004)\citenamefont
  {Anastasiou}, \citenamefont {Dixon}, \citenamefont {Melnikov},\ and\
  \citenamefont {Petriello}}]{Anastasiou:2003ds}%
  \BibitemOpen
  \bibfield  {author} {\bibinfo {author} {\bibfnamefont {C.}~\bibnamefont
  {Anastasiou}}, \bibinfo {author} {\bibfnamefont {L.~J.}\ \bibnamefont
  {Dixon}}, \bibinfo {author} {\bibfnamefont {K.}~\bibnamefont {Melnikov}},\
  and\ \bibinfo {author} {\bibfnamefont {F.}~\bibnamefont {Petriello}},\ }\href
  {https://doi.org/10.1103/PhysRevD.69.094008} {\bibfield  {journal} {\bibinfo
  {journal} {Phys. Rev. D}\ }\textbf {\bibinfo {volume} {69}},\ \bibinfo
  {pages} {094008} (\bibinfo {year} {2004})},\ \Eprint
  {https://arxiv.org/abs/hep-ph/0312266} {arXiv:hep-ph/0312266} \BibitemShut
  {NoStop}%
\bibitem [{\citenamefont {Becher}\ and\ \citenamefont
  {Neubert}(2011)}]{Becher:2010tm}%
  \BibitemOpen
  \bibfield  {author} {\bibinfo {author} {\bibfnamefont {T.}~\bibnamefont
  {Becher}}\ and\ \bibinfo {author} {\bibfnamefont {M.}~\bibnamefont
  {Neubert}},\ }\href {https://doi.org/10.1140/epjc/s10052-011-1665-7}
  {\bibfield  {journal} {\bibinfo  {journal} {Eur. Phys. J. C}\ }\textbf
  {\bibinfo {volume} {71}},\ \bibinfo {pages} {1665} (\bibinfo {year}
  {2011})},\ \Eprint {https://arxiv.org/abs/1007.4005} {arXiv:1007.4005
  [hep-ph]} \BibitemShut {NoStop}%
\bibitem [{\citenamefont {Becher}\ \emph {et~al.}(2008)\citenamefont {Becher},
  \citenamefont {Neubert},\ and\ \citenamefont {Xu}}]{Becher:2007ty}%
  \BibitemOpen
  \bibfield  {author} {\bibinfo {author} {\bibfnamefont {T.}~\bibnamefont
  {Becher}}, \bibinfo {author} {\bibfnamefont {M.}~\bibnamefont {Neubert}},\
  and\ \bibinfo {author} {\bibfnamefont {G.}~\bibnamefont {Xu}},\ }\href
  {https://doi.org/10.1088/1126-6708/2008/07/030} {\bibfield  {journal}
  {\bibinfo  {journal} {JHEP}\ }\textbf {\bibinfo {volume} {07}},\ \bibinfo
  {pages} {030}},\ \Eprint {https://arxiv.org/abs/0710.0680} {arXiv:0710.0680
  [hep-ph]} \BibitemShut {NoStop}%
\bibitem [{\citenamefont {Gribov}\ and\ \citenamefont
  {Lipatov}(1972)}]{Gribov:1972rt}%
  \BibitemOpen
  \bibfield  {author} {\bibinfo {author} {\bibfnamefont {V.~N.}\ \bibnamefont
  {Gribov}}\ and\ \bibinfo {author} {\bibfnamefont {L.~N.}\ \bibnamefont
  {Lipatov}},\ }\href@noop {} {\bibfield  {journal} {\bibinfo  {journal} {Sov.
  J. Nucl. Phys.}\ }\textbf {\bibinfo {volume} {15}},\ \bibinfo {pages} {675}
  (\bibinfo {year} {1972})}\BibitemShut {NoStop}%
\bibitem [{\citenamefont {Dokshitzer}(1977)}]{Dokshitzer:1977sg}%
  \BibitemOpen
  \bibfield  {author} {\bibinfo {author} {\bibfnamefont {Y.~L.}\ \bibnamefont
  {Dokshitzer}},\ }\href@noop {} {\bibfield  {journal} {\bibinfo  {journal}
  {Sov. Phys. JETP}\ }\textbf {\bibinfo {volume} {46}},\ \bibinfo {pages} {641}
  (\bibinfo {year} {1977})}\BibitemShut {NoStop}%
\bibitem [{\citenamefont {Altarelli}\ and\ \citenamefont
  {Parisi}(1977)}]{Altarelli:1977zs}%
  \BibitemOpen
  \bibfield  {author} {\bibinfo {author} {\bibfnamefont {G.}~\bibnamefont
  {Altarelli}}\ and\ \bibinfo {author} {\bibfnamefont {G.}~\bibnamefont
  {Parisi}},\ }\href {https://doi.org/10.1016/0550-3213(77)90384-4} {\bibfield
  {journal} {\bibinfo  {journal} {Nucl. Phys. B}\ }\textbf {\bibinfo {volume}
  {126}},\ \bibinfo {pages} {298} (\bibinfo {year} {1977})}\BibitemShut
  {NoStop}%
\bibitem [{\citenamefont {Lepage}(1978)}]{Lepage:1977sw}%
  \BibitemOpen
  \bibfield  {author} {\bibinfo {author} {\bibfnamefont {G.~P.}\ \bibnamefont
  {Lepage}},\ }\href {https://doi.org/10.1016/0021-9991(78)90004-9} {\bibfield
  {journal} {\bibinfo  {journal} {J. Comput. Phys.}\ }\textbf {\bibinfo
  {volume} {27}},\ \bibinfo {pages} {192} (\bibinfo {year} {1978})}\BibitemShut
  {NoStop}%
\bibitem [{\citenamefont {Lepage}(2021)}]{Lepage:2020tgj}%
  \BibitemOpen
  \bibfield  {author} {\bibinfo {author} {\bibfnamefont {G.~P.}\ \bibnamefont
  {Lepage}},\ }\href {https://doi.org/10.1016/j.jcp.2021.110386} {\bibfield
  {journal} {\bibinfo  {journal} {J. Comput. Phys.}\ }\textbf {\bibinfo
  {volume} {439}},\ \bibinfo {pages} {110386} (\bibinfo {year} {2021})},\
  \Eprint {https://arxiv.org/abs/2009.05112} {arXiv:2009.05112
  [physics.comp-ph]} \BibitemShut {NoStop}%
\bibitem [{\citenamefont {Gaunt}(2014)}]{Gaunt:2014ska}%
  \BibitemOpen
  \bibfield  {author} {\bibinfo {author} {\bibfnamefont {J.~R.}\ \bibnamefont
  {Gaunt}},\ }\href {https://doi.org/10.1007/JHEP07(2014)110} {\bibfield
  {journal} {\bibinfo  {journal} {JHEP}\ }\textbf {\bibinfo {volume} {07}},\
  \bibinfo {pages} {110}},\ \Eprint {https://arxiv.org/abs/1405.2080}
  {arXiv:1405.2080 [hep-ph]} \BibitemShut {NoStop}%
\bibitem [{\citenamefont {Zeng}(2015)}]{Zeng:2015iba}%
  \BibitemOpen
  \bibfield  {author} {\bibinfo {author} {\bibfnamefont {M.}~\bibnamefont
  {Zeng}},\ }\href {https://doi.org/10.1007/JHEP10(2015)189} {\bibfield
  {journal} {\bibinfo  {journal} {JHEP}\ }\textbf {\bibinfo {volume} {10}},\
  \bibinfo {pages} {189}},\ \Eprint {https://arxiv.org/abs/1507.01652}
  {arXiv:1507.01652 [hep-ph]} \BibitemShut {NoStop}%
\bibitem [{\citenamefont {Buckley}\ \emph {et~al.}(2015)\citenamefont
  {Buckley}, \citenamefont {Ferrando}, \citenamefont {Lloyd}, \citenamefont
  {Nordstr\"om}, \citenamefont {Page}, \citenamefont {R\"ufenacht},
  \citenamefont {Sch\"onherr},\ and\ \citenamefont {Watt}}]{Buckley:2014ana}%
  \BibitemOpen
  \bibfield  {author} {\bibinfo {author} {\bibfnamefont {A.}~\bibnamefont
  {Buckley}}, \bibinfo {author} {\bibfnamefont {J.}~\bibnamefont {Ferrando}},
  \bibinfo {author} {\bibfnamefont {S.}~\bibnamefont {Lloyd}}, \bibinfo
  {author} {\bibfnamefont {K.}~\bibnamefont {Nordstr\"om}}, \bibinfo {author}
  {\bibfnamefont {B.}~\bibnamefont {Page}}, \bibinfo {author} {\bibfnamefont
  {M.}~\bibnamefont {R\"ufenacht}}, \bibinfo {author} {\bibfnamefont
  {M.}~\bibnamefont {Sch\"onherr}},\ and\ \bibinfo {author} {\bibfnamefont
  {G.}~\bibnamefont {Watt}},\ }\href
  {https://doi.org/10.1140/epjc/s10052-015-3318-8} {\bibfield  {journal}
  {\bibinfo  {journal} {Eur. Phys. J. C}\ }\textbf {\bibinfo {volume} {75}},\
  \bibinfo {pages} {132} (\bibinfo {year} {2015})},\ \Eprint
  {https://arxiv.org/abs/1412.7420} {arXiv:1412.7420 [hep-ph]} \BibitemShut
  {NoStop}%
\bibitem [{\citenamefont {Forshaw}\ \emph {et~al.}(2020)\citenamefont
  {Forshaw}, \citenamefont {Holguin},\ and\ \citenamefont
  {Pl\"atzer}}]{Forshaw:2020wrq}%
  \BibitemOpen
  \bibfield  {author} {\bibinfo {author} {\bibfnamefont {J.~R.}\ \bibnamefont
  {Forshaw}}, \bibinfo {author} {\bibfnamefont {J.}~\bibnamefont {Holguin}},\
  and\ \bibinfo {author} {\bibfnamefont {S.}~\bibnamefont {Pl\"atzer}},\ }\href
  {https://doi.org/10.1007/JHEP09(2020)014} {\bibfield  {journal} {\bibinfo
  {journal} {JHEP}\ }\textbf {\bibinfo {volume} {09}},\ \bibinfo {pages}
  {014}},\ \Eprint {https://arxiv.org/abs/2003.06400} {arXiv:2003.06400
  [hep-ph]} \BibitemShut {NoStop}%
\bibitem [{\citenamefont {Nagy}\ and\ \citenamefont
  {Soper}(2020)}]{Nagy:2020dvz}%
  \BibitemOpen
  \bibfield  {author} {\bibinfo {author} {\bibfnamefont {Z.}~\bibnamefont
  {Nagy}}\ and\ \bibinfo {author} {\bibfnamefont {D.~E.}\ \bibnamefont
  {Soper}},\ }\href@noop {} {\  (\bibinfo {year} {2020})},\ \Eprint
  {https://arxiv.org/abs/2011.04777} {arXiv:2011.04777 [hep-ph]} \BibitemShut
  {NoStop}%
\bibitem [{\citenamefont {Herren}\ \emph {et~al.}(2022)\citenamefont {Herren},
  \citenamefont {H\"oche}, \citenamefont {Krauss}, \citenamefont {Reichelt},\
  and\ \citenamefont {Schoenherr}}]{Herren:2022jej}%
  \BibitemOpen
  \bibfield  {author} {\bibinfo {author} {\bibfnamefont {F.}~\bibnamefont
  {Herren}}, \bibinfo {author} {\bibfnamefont {S.}~\bibnamefont {H\"oche}},
  \bibinfo {author} {\bibfnamefont {F.}~\bibnamefont {Krauss}}, \bibinfo
  {author} {\bibfnamefont {D.}~\bibnamefont {Reichelt}},\ and\ \bibinfo
  {author} {\bibfnamefont {M.}~\bibnamefont {Schoenherr}},\ }\href@noop {} {\
  (\bibinfo {year} {2022})},\ \Eprint {https://arxiv.org/abs/2208.06057}
  {arXiv:2208.06057 [hep-ph]} \BibitemShut {NoStop}%
\bibitem [{\citenamefont {Becher}\ \emph {et~al.}(2023)\citenamefont {Becher},
  \citenamefont {Favrod},\ and\ \citenamefont {Xu}}]{Becher:2022rhu}%
  \BibitemOpen
  \bibfield  {author} {\bibinfo {author} {\bibfnamefont {T.}~\bibnamefont
  {Becher}}, \bibinfo {author} {\bibfnamefont {S.}~\bibnamefont {Favrod}},\
  and\ \bibinfo {author} {\bibfnamefont {X.}~\bibnamefont {Xu}},\ }\href
  {https://doi.org/10.1007/JHEP01(2023)005} {\bibfield  {journal} {\bibinfo
  {journal} {JHEP}\ }\textbf {\bibinfo {volume} {01}},\ \bibinfo {pages}
  {005}},\ \Eprint {https://arxiv.org/abs/2208.01554} {arXiv:2208.01554
  [hep-ph]} \BibitemShut {NoStop}%
\bibitem [{\citenamefont {Ferrario~Ravasio}\ \emph {et~al.}(2023)\citenamefont
  {Ferrario~Ravasio}, \citenamefont {Hamilton}, \citenamefont {Karlberg},
  \citenamefont {Salam}, \citenamefont {Scyboz},\ and\ \citenamefont
  {Soyez}}]{FerrarioRavasio:2023kyg}%
  \BibitemOpen
  \bibfield  {author} {\bibinfo {author} {\bibfnamefont {S.}~\bibnamefont
  {Ferrario~Ravasio}}, \bibinfo {author} {\bibfnamefont {K.}~\bibnamefont
  {Hamilton}}, \bibinfo {author} {\bibfnamefont {A.}~\bibnamefont {Karlberg}},
  \bibinfo {author} {\bibfnamefont {G.~P.}\ \bibnamefont {Salam}}, \bibinfo
  {author} {\bibfnamefont {L.}~\bibnamefont {Scyboz}},\ and\ \bibinfo {author}
  {\bibfnamefont {G.}~\bibnamefont {Soyez}},\ }\href
  {https://doi.org/10.1103/PhysRevLett.131.161906} {\bibfield  {journal}
  {\bibinfo  {journal} {Phys. Rev. Lett.}\ }\textbf {\bibinfo {volume} {131}},\
  \bibinfo {pages} {161906} (\bibinfo {year} {2023})},\ \Eprint
  {https://arxiv.org/abs/2307.11142} {arXiv:2307.11142 [hep-ph]} \BibitemShut
  {NoStop}%
\end{thebibliography}%


\onecolumngrid
\newpage
\appendix

\section*{Supplemental material}
\thispagestyle{empty}
\setcounter{page}{1}

In the following, we list the ingredients for the NLL parton shower and provide details on their technical implementation. We first provide the angular functions in the two-loop anomalous dimension, and then give the order $\alpha_s$ hard functions for $pp \to Z$ and explain their implementation. Next, we specify the MC algorithm for computing the contribution of the two-loop anomalous dimension. In the final two sections, we show plots comparing the $N_c=3$ result to the large-$N_c$ approximation to the LL result and provide some details on the comparison to {\sc Gnole}.

\subsection{A. Angular functions in the anomalous dimension}
\renewcommand{\theequation}{A.\arabic{equation}}
\setcounter{equation}{0}
\label{app:anfcts}
The two-loop anomalous dimension of \cite{Becher:2021urs} presented in the main text involves some angular functions, as well as the two-loop cusp anomalous dimension
\begin{align}
\gamma_1^\text{cusp} = 4 \left(\left(\frac{67}{9}-\frac{\pi ^2}{3}\right) C_A-\frac{20}{9}  n_F T_F \right) , 
\end{align}
with $n_F$ the number of fermions included in the theory and $T_F$ the trace of the associated generator. The one- and two-loop $\beta$-function coefficients are 
\begin{align}
\beta_0&=\frac{11}{3} \spac C_A-\frac{4}{3} \spac n_F\spac T_F\,, &
\beta_1 &= \frac{34}{3}C_A^{\spac 2} - \frac{20}{3}C_AT_F \spac n_F-4\spac C_FT_F \spac n_F\,.
\end{align}
In the strict large-$N_c$ limit we could drop the $n_F$ terms, but we will keep these contributions, both in the $\beta$-function and in the anomalous dimension $\bm{\Gamma}^{(2)}$. The $\beta_0$-terms in $\bm{d}_m$ and $\bm{r}_m$ involve the angular function
\begin{align}
X_{ij}^q = W_{ij}^q \cdot \ln(4 s_q^2)\,,
\end{align}
where $s_q = \sin(\phi_{q})$ is the sine of the azimuthal angle of the soft emission $n_q$ in the frame where the directions $i$ and $j$ are back-to-back. This term arises from the combination 
\begin{equation}\label{eq:comb}
   \int \! \left[d^2\Omega_{q}\right]  \ln(2W_{ij}^q) \, W_{ij}^q -2  \!\int \! \left[d\Omega_{q} \right]_{\epsilon} W_{ij}^q  = \int \! \left[d^2\Omega_{q}\right] \, X_{ij}^q\,,
\end{equation}
where the angular integration in $d=4-2\epsilon$ dimension was expanded as
 \begin{equation}
 \int [d \Omega_{q}] =  \int \! \left[d\Omega_{q}\right]_{2}  + 2\epsilon  \!\int \! \left[d\Omega_{q} \right]_{\epsilon} + \mathcal{O}(\epsilon^2)\,.
 \end{equation}
 The $\ln(2W_{ij}^q)$ term in \eqref{eq:comb} is present in the diagrammatic result for the anomalous dimension with angular integrals defined in $d=4-2\epsilon$. The order $\epsilon$ terms in the angular integrations arise when converting the original result for $\bm{\Gamma}^{(2)}$ to the $\overline{\text{MS}}$ scheme in which the angular integrals are defined in $d=4$, see the discussion in Section 5.3 in \cite{Becher:2021urs}. The two terms in \eqref{eq:comb} are not separately Lorentz invariant but their combination is. The lack of Lorentz invariance of the diagrammatic result was noticed in \cite{CaronHuot:2015bja}. The fact that the anomalous dimension is Lorentz invariant up to the angular constraints is useful for its implementation since it allows us to evaluate the angular integrals in an arbitrary frame. A detailed discussion of the invariance properties of angular integrals will be presented elsewhere \cite{BecherSchalchXuUpcoming}.

A second Lorentz invariant combination arises from combining the scheme conversion commutator terms derived in Section 5.3 in \cite{Becher:2021urs} and the three-leg terms involving the angular function $K_{iij;qr}+K_{jji;qr}$ introduced in \cite{CaronHuot:2015bja}. 
Separately, these two are not Lorentz invariant, but together they yield the structure \cite{BecherSchalchXuUpcoming}
\begin{align}
M_{ij;qr}&= \left(W_{ij}^q W_{ij}^r -W_{ij}^{qr}-W_{ij}^{rq}\right) \ln \frac{s_{qr}^2}{s_q^2}\, , 
\label{eq:stronglyordered}
\end{align}
with $W_{ij}^{qr}= W_{ij}^q W_{qj}^r$. The quantity $M_{ij;qr}$ corresponds to the strongly ordered emissions multiplied by the logarithm of the ratio between $s_{qr}^2=\sin^2(\phi _{qr})$, the sine squared of the azimuthal angle difference $\phi _{qr}=\phi_q-\phi_r$ and $s_q^2 = \sin^2(\phi_q)$ in the frame where $i$ and $j$ are back-to-back. This quantity can be written in a manifestly Lorentz invariant way as
\begin{align}
4s_{qr}^2=&\frac{1}{n_{iq}\spac n_{ir} \spac n_{jq} \spac n_{jr}}\Big[2 \spac n_{ij} \spac n_{qr} \left(n_{ir} \spac n_{jq}+n_{iq}\spac n_{jr}\right) -\left(n_{ij}\right){}^2 \left(n_{qr}\right){}^2-\left(n_{ir} \spac n_{jq}-n_{iq} \spac n_{jr}\right){}^2   \Big]\,.
\end{align}
In addition, we need the two-particle terms $K_{ij;qr}\spac$, which are split up as \cite{CaronHuot:2015bja,Becher:2021urs}
\begin{align}\label{eq:Kijqr}
 K_{ij;qr} = C_A K_{ij;qr}^{(a)} + \left[ n_F T_F-2C_A \right] K_{ij;qr}^{(b)} + \left[C_A -2n_FT_F\right] K_{ij;qr}^{(c)}\,.
 \end{align}
The individual functions are specified by
\begin{equation}
\begin{aligned} \label{eq:KijqrFuns}
K_{ij;qr}^{(a)} &= \frac{4n_{ij}}{n_{iq}n_{qr}n_{jr}}\left[1+\frac{n_{ij}n_{qr}}{n_{iq}n_{jr}-n_{ir}n_{jq}}\right]\ln\frac{n_{iq}n_{jr}}{n_{ir}n_{jq}} \,, \\
K_{ij;qr}^{(b)} &= \frac{8 n_{ij}}{n_{qr}(n_{iq}n_{jr}-n_{ir}n_{jq})}\ln\frac{n_{iq}n_{jr}}{n_{ir}n_{jq}} \, ,\\
K_{ij;qr}^{(c)} & =\frac{4}{n_{qr}^2}\,\left( \frac{n_{iq}n_{jr}+n_{ir}n_{jq}}{n_{iq}n_{jr}-n_{ir}n_{jq}}\ln\frac{n_{iq}n_{jr}}{n_{ir}n_{jq}} -2 \right)  . 
\end{aligned}
\end{equation}
These functions are finite when $n_q$ or $n_r$ become collinear to the legs $n_i$ and $n_j$, however, collinear divergences arise when $n_q$ becomes collinear to $n_r$. The terms in the double-emission contribution $\boldsymbol{d}_m$ proportional to $n_F$ describe the splitting of a gluon into a quark-antiquark pair. The large-$N_c$ structure of this contribution differs from the remaining terms describing the emission of two gluons. To account for this in our shower, the insertion of $\boldsymbol{d}_m$ is split up according to 
\begin{equation}
  \boldsymbol{d}^{\spac ij}_m \hspace{2mm}
  \begin{tikzpicture}[baseline={([yshift=-.5ex]current bounding box.center)},vertex/.style={anchor=base,
    circle,fill=black!25,minimum size=18pt,inner sep=2pt}]
    \draw[line width=0.25mm] (1,1) -- (2,2) ;
    \draw[line width=0.25mm] (1,1) -- (2,0) ;
    \node at (2.3,2){\scalebox{0.8}{$n_i$}};
    \node at (2.3,-0.05){\scalebox{0.8}{$n_j$}};
  \end{tikzpicture}
   \hspace{2mm} = \hspace{5mm}
   \begin{tikzpicture}[baseline={([yshift=-.5ex]current bounding box.center)},vertex/.style={anchor=base,
    circle,fill=black!25,minimum size=18pt,inner sep=2pt}]
    \draw[line width=0.25mm] (1,1.5) -- (2,2.5) ;
    \draw[line width=0.25mm] (1,1.5) -- (2,1.5) ;
    \draw[line width=0.25mm] (1,1.35) -- (2,1.35) ;
    \draw[line width=0.25mm] (1,1.35) -- (1,0.65) ;
    \draw[line width=0.25mm] (1,0.65) -- (2,0.65) ;
    \draw[line width=0.25mm] (1,0.5) -- (2,0.5) ;
    \draw[line width=0.25mm] (1,0.5) -- (2,-0.5) ;
    \node at (2.3,2.5){\scalebox{0.8}{$n_i$}};
    \node at (2.3,1.4){\scalebox{0.8}{$n_q$}};
    \node at (2.3,0.55){\scalebox{0.8}{$n_r$}};
    \node at (2.3,-0.55){\scalebox{0.8}{$n_j$}};
  \end{tikzpicture} 
  + \hspace{2mm} n_F \hspace{2mm}
   \begin{tikzpicture}[baseline={([yshift=-.5ex]current bounding box.center)},vertex/.style={anchor=base,
    circle,fill=black!25,minimum size=18pt,inner sep=2pt}]
    \draw[line width=0.25mm] (1,1.2) -- (2,2.2) ;
    \draw[line width=0.25mm] (1,1.2) -- (2,1.2) ;
    \draw[line width=0.25mm] (1,0.8) -- (2,0.8) ;
    \draw[line width=0.25mm] (1,0.8) -- (2,-0.2) ;
    \node at (2.3,2.2){\scalebox{0.8}{$n_i$}};
    \node at (2.3,1.2){\scalebox{0.8}{$n_q$}};
    \node at (2.3,0.7){\scalebox{0.8}{$n_r$}};
    \node at (2.3,-0.25){\scalebox{0.8}{$n_j$}};
  \end{tikzpicture}
  ,
\end{equation}
where the lines indicate the dipole structure after the insertion. The evolution after the insertion in \eqref{eq:2looprunning} is then performed separately for the two parts, according to their dipole structure. On a technical level, it is convenient to treat $n_q$ and $n_r$ as an additional dipole with $V_{qr}=0$ in the $n_F$ part, i.e., as a dipole which never radiates.
\subsection{B. Hard matching corrections for \texorpdfstring{$\bm{Z}$}{Z} production}
\renewcommand{\theequation}{B.\arabic{equation}}
\setcounter{equation}{0}
In this section we derive the radiative corrections to the hard functions $\bm{\mathcal{H}}_3^{(1)},\bm{\mathcal{H}}_2^{(1)}$ which can be obtained by a standard next-to-leading order (NLO) QCD calculation \cite{Altarelli:1979ub,Anastasiou:2003yy,Anastasiou:2003ds} for the Drell-Yan process \cite{Drell:1970wh}. The production of a (virtual) color-singlet boson with invariant mass $Q$ in hadron-hadron collisions at center-of-mass energy $\sqrt{s}$ occurs through the annihilation of quarks. More precisely, we study the process $N_1+N_2\rightarrow Z+ X_{\rm had}$, where $N_1$ and $N_2$ are the colliding hadrons (protons at the LHC) and $X_{\rm had}$ is an arbitrary hadronic final state. The cross section can be expressed as a convolution of the partonic cross sections $\frac{d\hat{\sigma}_{ij}}{dQ^2}$ with parton distribution functions\footnote{Non-global hadron-collider cross sections at finite $N_c$ involve super-leading logarithms due to Glauber phases \cite{Forshaw:2006fk,Becher:2021zkk}. One might wonder whether the Glauber effects would lead to a breakdown of PDF factorization. Indeed, the papers \cite{Gaunt:2014ska,Zeng:2015iba} suggest that even global observables (e.g.\ jet vetoes) would suffer from factorization-breaking effects due to Glauber gluons. Whether PDF factorization holds beyond inclusive Drell-Yan is an important open question for all of hadron collider physics, but beyond the scope of the present work.} $f_{i/N}(x,\mu_f)$ (PDFs)
\begin{align}
\frac{d \sigma}{d Q^2} = \sum_{i,j=q,\bar{q},g}\int dx_1\int dx_2 \spac \spac f_{i/N_1}(x_1,\mu_f)\spac f_{j/N_2}(x_2,\mu_f)\,  \sigma^{q}_0 \,\frac{d\hat{\sigma}_{ij}}{dQ^2}\,.\label{eq:hadrxsec}
\end{align}
For convenience, we have introduced a prefactor
\begin{equation}
\sigma^{q}_0 = \frac{4\pi^2 \alpha}{N_c \spac s}\sum_q\frac{|g_L^q|^2+|g_R^q|^2}{2}\,,
\end{equation}
which absorbs the electromagnetic coupling  $\alpha$ and the left- and right-handed couplings $g_{L,R}$ of the $Z$ to fermions. At NLO, each partonic channel only involves a single (anti-)quark flavor $q$. The explicit form of the couplings for different electroweak bosons can be found in \cite{Becher:2010tm}; for simplicity, we only consider $Z$ production in the present work. The PDFs $f_{i/N}(x,\mu_f)$ are related to the probability of finding a parton $i$ inside the hadron $N$ with momentum fraction $x$, and $\mu_f$ is the factorization scale. In the center-of-mass frame, we parametrize the incoming hadron-momenta as $P_{1,2}=\frac{\sqrt{s}}{2}(1,0,0,\pm 1)$ which translates to, as usual, $p_1=x_1P_1$ and $p_2=x_2 P_2$ at the partonic level. We find it useful to introduce the variables (see e.g. \cite{Becher:2007ty})
\begin{align}
\tau = \frac{Q^2}{s}\,, \hspace*{5mm} z = \frac{Q^2}{\hat{s}} = \frac{\tau}{x_1x_2}\,,
\end{align}
where $\hat{s}=(p_1+p_2)^2=x_1x_2\spac s$ denotes the center-of-mass energy squared in the partonic system. In addition, we also introduce the Mandelstam variable $\hat{t}=(p_1-k)^2$, where $k$ is the momentum of the final-state gluon or quark at NLO.
Factoring out the electroweak prefactors as in \eqref{eq:hadrxsec}, the partonic cross section can be written in terms of hard scattering kernels $C_{ij}$
\begin{align}
\frac{d\hat{\sigma}_{ij}}{dQ^2}= \int \mathrm{d}\Pi_f \spac\spac |\mathcal{M}_{ij}|^2\spac\spac\delta\Big(z-\tfrac{Q^2}{\hat{s}}\Big)=\int_\tau^1 dz \int_0^1 dy \spac\spac\spac C_{ij}(z,y,Q,\mu_f) \spac \spac \delta\Big(z-\tfrac{Q^2}{\hat{s}}\Big)\,,
\end{align}
with $y = 1+\frac{\hat{t}}{\hat{s}+Q^2}$, which is related to the scattering angle~$\theta$ in the partonic center-of-mass frame through $\cos \theta = 2y-1$. We have written the phase space of the final-state particles $\mathrm{d}\Pi_f$ in terms of $y,z$ and $Q$. The hard scattering kernels $C_{ij}$ are expanded perturbatively in $\alpha_s$. At leading order, i.e. $\alpha_s^0$, only the $q\spac \overline{q}$ channel contributes, whereas at NLO the inelastic $qg$ channel opens up (of course, together with the directly related $\overline{q}q$, $\bar{q}g$, $gq$ and $g\bar{q}$ channels). After combining real and virtual corrections and performing the collinear factorization, we obtain the LO and NLO hard scattering kernels in the $\overline{\mathrm{MS}}$ scheme
\begin{align}
    C_{q\overline{q}} = & \spac\spac\spac \delta(1-z) \spac \tfrac{\delta(y)+\delta(1-y)}{2}\Bigg[1+\frac{\alpha_s}{4\pi}\hspace{1mm} C_F\bigg(8\zeta_2-16-6\ln \tfrac{\mu_f^2}{Q^2}\bigg)\Bigg]\nonumber\\[2mm]
    \ &  +\frac{\alpha_s}{4\pi}\hspace{1mm} C_F  \Bigg[2\big(\delta(y)+\delta(1-y)\big)\Bigg(4\Big[\tfrac{\ln(1-z)}{1-z}\Big]_+ -2\hspace{0.5mm}(1+z)\ln(1-z) - \tfrac{(1+z^2)}{1-z}\ln(z) \nonumber\\[1mm]
   & \hspace{4.1cm} +1 -z - (1+z^2)\Big[\tfrac{1}{1-z}\Big]_+\ln \tfrac{\mu_f^2}{Q^2}\Bigg)  \nonumber\\[1mm]
   & \hspace{2cm} + 2\hspace{1mm} \bigg((1+z^2)\Big[\tfrac{1}{1-z}\Big]_+\bigg(\Big[\tfrac{1}{y}\Big]_+ + \Big[\tfrac{1}{1-y}\Big]_+\bigg)-2\hspace{0.5mm}(1-z)\bigg) \Bigg]\nonumber\\[2mm]
  \ &+\frac{\alpha_s}{4\pi}\hspace{1mm} C_F \spac\spac\spac \delta(1-z) \Bigg[2\zeta_2\spac\spac\big(\delta(y)+\delta(1-y)\big)+2\bigg(\tfrac{\ln (1-y)}{y}+\tfrac{\ln (y)}{1-y}+\Big[\tfrac{\ln (y)}{y}\Big]_+ + \Big[\tfrac{\ln (1-y)}{1-y}\Big]_+\bigg) \nonumber\\[2mm]
  & \hspace*{3cm}-2\ln \tfrac{\mu_h^2}{Q^2} \bigg(\Big[\tfrac{1}{y}\Big]_+ + \Big[\tfrac{1}{1-y}\Big]_+\bigg)\Bigg]\, ,\label{eq:nloqq}\\[4mm]
  C_{qg} =   & \hspace{1mm} \frac{\alpha_s}{4 \pi} \hspace{1mm} T_F \hspace{0.5mm}  \Bigg[2 \hspace{1mm} \delta(y)\hspace{1mm}\bigg(\big(z^2+(1-z)^2\big)\ln \Big(\tfrac{(1-z)^2}{z}\Big) + 2 z(1-z) -\big(z^2+(1-z)^2\big)\ln \tfrac{\mu_f^2}{Q^2}\bigg)\nonumber \\
   & \hspace{13mm} + 2\hspace{1mm} \bigg(\big(z^2+(1-z)^2\big)\Big[\tfrac{1}{y}\Big]_+ +2z(1-z) +(1-z)^2 \hspace{0.5mm} y\bigg) \Bigg]\, .\label{eq:nloqg}
\end{align}
The remaining channels can be reconstructed from these results. Using the fact that \eqref{eq:hadrxsec} is scale independent, we may derive the dependence on $\mu_f$ in \eqref{eq:nloqq} and \eqref{eq:nloqg}, using the fact that the PDFs obey the DGLAP evolution equations \cite{Gribov:1972rt,Dokshitzer:1977sg,Altarelli:1977zs}. As pointed out in the main text, the expressions for partonic cross sections are usually simplified by integrating soft contributions over all angles. In our case, this concerns the terms in the last two lines of \eqref{eq:nloqq}, which can be written in compact form as
\begin{align}\label{eq:extra}
\Delta \hat{\sigma}_{q\bar{q}} = 
\frac{C_F\spac \alpha_s}{2\pi} \spac\delta(1-z) \left\{ \left[\tfrac{\ln\left((1-y)\spac y\right)}{(1-y) y}\right]_+  - \ln\tfrac{\mu_h^2}{Q^2} \left[\tfrac{1}{(1-y)\spac y}\right]_+ \right\}\,.
\end{align}
These terms were omitted in \cite{Anastasiou:2003yy} since they integrate to zero in perturbative predictions for infrared safe observables. However, we use the partonic cross section to construct the hard functions \eqref{eq:hardfct} in the factorization theorem and due to the restriction $\Theta_{\rm in}\!\left(\left\{\underline{n}\right\}\right)$ in \eqref{eq:hardfct} also the $\delta(1-z)$ terms are only integrated over the jet region. The part of the cross section where the gluon is inside the gap is included in the soft function and integrating the terms \eqref{eq:extra} in the hard function over the full angle would lead to a double counting. The terms in the last line of \eqref{eq:nloqq} depend on the renormalization scale which we denote by $\mu_h$, since the above partonic cross sections will be part of the hard function. For the same reason,  the coupling $\alpha_s\equiv \alpha_s(\mu_h)$ is evaluated at this scale.

We now want to use the results \eqref{eq:nloqq} and \eqref{eq:nloqg} for the calculation of the angular convolutions of the hard and soft functions. For the hard matching corrections, we need the one-loop corrections to the hard function $\bm{\mathcal{H}}_2$, together with the leading-order hard function $\bm{\mathcal{H}}_3$ in the combination
\begin{align}
\frac{\alpha_s}{4\pi}\Delta^{(1)} = \frac{\alpha_s} {4\pi} \big\langle\bm{\mathcal{H}}_2^{(1)}\otimes \bm{\mathcal{S}}^{\rm LL}_2(t)+ \bm{\mathcal{H}}_3^{(1)}\otimes \bm{\mathcal{S}}^{\rm LL}_3(t) \big\rangle\,, \label{eq:hardmatch}
\end{align}
where the LL soft functions include the evolution from the hard to the soft scale
\begin{equation}
\bm{\mathcal{S}}^{\rm LL}_m(t) = \sum_k \bm{U}_{mk}(t_0=0,t)\spac\hat{\otimes}\spac \bm{1}\,,
\end{equation}
and the leading-order soft functions are trivial. The symbol $\hat{\otimes}$ indicates the integration over the directions of the additional partons generated during the evolution. 

For the hard function $\bm{\mathcal{H}}_2$, the angular convolution in \eqref{eq:hardmatch} is  trivial since the two incoming partons are along the beam directions. For the function $\bm{\mathcal{H}}_3$, a single nontrivial angular integral remains, corresponding to the angle of the emitted final-state particle with respect to the beam axis. We parameterize this angle in the variable $y$ so that we have
\begin{equation}
\begin{aligned}
 \Delta^{(1)} &=\,\big\langle\bm{\mathcal{H}}_2^{(1)}(z) \,\bm{\mathcal{S}}_2(t)+ \int_0^1 dy\, \bm{\mathcal{H}}_3^{(1)}(y,z)\, \bm{\mathcal{S}}_3(y,t) \big\rangle \\
&= \int_0^1 dy \spac\spac \Big[\tfrac{\delta (y)+\delta (1-y)}{2}\spac\spac\bm{\mathcal{H}}_{2}^{(1)}\spac (z) +\bm{\mathcal{H}}_{3}^{(1)}(y,z) \Big] \spac\bm{\mathcal{S}}_3(y,t)\,,
\end{aligned}\label{eq:hardmatchComb}
\end{equation}
where we use that the soft function $\bm{\mathcal{S}}_3(y,t)$ reduces to the two-parton soft function when the emitted parton is along the beam direction, i.e. $\bm{\mathcal{S}}_3(1,t)=\bm{\mathcal{S}}_3(0,t)=\bm{\mathcal{S}}_2(t)$. The soft functions $\bm{\mathcal{S}}_3(y,t)$ are computed by starting the LL shower with an appropriate three-parton configuration. 

The advantage of combining the real and virtual corrections in \eqref{eq:hardmatchComb} is that the combined hard functions are directly related to the partonic amplitudes given above. For example, for the $q\bar{q}$ channel, we have
\begin{equation}
\tfrac{\delta (y)+\delta (1-y)}{2}\spac\spac\bm{\mathcal{H}}_{2,q\bar{q}}\spac (z) +\bm{\mathcal{H}}_{3,q\bar{q}}(y,z) = \sigma_0^q\, C_{q\bar{q}}(z,y)\, \Theta_{\rm in}(y,z)\,,
\end{equation}
where we express the in-jet constraint $\Theta_{\rm in}(y,z)$ through the partonic variables $y$ and $z$. For a gap around the rapidity of the $Z$-boson, we have
\begin{equation}
\Theta_{\rm in}(y,z) = \theta\Big(|Y| - Y_{\rm max}\Big)\,,
\end{equation}
where 
\begin{align}
Y&= \frac{1}{2}\ln \frac{\hat{y}}{1-\hat{y}}\,, & &\text{ with }& \hat{y} &= \frac{y \spac (y \spac (1-z)+z)}{1-2 \spac (1-y) \spac y \spac (1-z)}\,.
\end{align}
The quark-gluon channel is not present at the lowest order, so that
\begin{equation}
\bm{\mathcal{H}}_{3,qg}(y,z) = \sigma^q_0 \,C_{qg}(z,y)\, \Theta_{\rm in}(y,z)\,.
\end{equation}

In our code, we first calculate the soft functions $\spac\bm{\mathcal{S}}_3^{\spac qg}(y,t),\bm{\mathcal{S}}_3^{\spac q\overline{q}}(y,t)$ numerically via LL evolution for a grid of $y$ values. We then interpolate in $y$ and evaluate the convolution with the partonic amplitudes. The integrations in $y,z$ as well as $x_i$ are performed using VEGAS \cite{Lepage:1977sw,Lepage:2020tgj}. To simplify the matching to fixed order, we compute the order $\alpha_s$ corrections to the hard and soft functions for $N_c=3$.

\subsection{C. Shower algorithm}\label{sec:showeralgo}
\renewcommand{\theequation}{C.\arabic{equation}}
\setcounter{equation}{0}
\begin{figure}
\includegraphics[scale=0.8,valign=t]{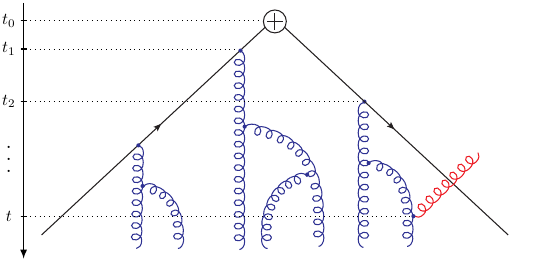}\hspace{20mm}
\includegraphics[scale=0.8,valign=t]{./figures/NLL_shower}
\caption{\label{fig:LLshower} Pictorial representations of the LL shower (left) and the NLL shower (right). Blue lines denote hard emissions inside the jets generated by the shower evolution. The red line depicts a soft emission into the veto region, which terminates the shower. The pink blob is an insertion of the two-loop double-real contribution $\bm{d}_m$ at time~$t'$.} 
\end{figure}
In the following, we describe the NLL shower algorithm. The LL shower was described in detail in \cite{Balsiger:2018ezi,Balsiger:2020ogy} and is closely related to the one introduced in \cite{Dasgupta:2001sh}. Reference \cite{Balsiger:2019tne} detailed the computation of the one-loop soft function and the implementation of $\bm{\mathcal{H}}_3$ into the shower. We will assume that the reader is familiar with the LL shower and will explain the computation of 
\begin{align}
\Delta\bm{U}_{kl}(t_0,t)=\int_{t_0}^t \! dt' \spac\spac\spac \bm{U}_{kk'}(t_0,t')\cdot \frac{\alpha_s(t')}{4\pi}\spac\Big(\bm{\Gamma}^{(2)}_{k'l'}-\frac{\beta_1}{\beta_0} \bm{\Gamma}^{(1)}_{k'l'}\Big)\cdot \bm{U}_{l'l}(t',t).\label{eq:2looprunningApp}
\end{align}
A fundamental ingredient of our shower is a list of $m$ vectors $\{\underline{n}\}= \{n_1,n_{i_1}, \dots,n_{i_{m-2}},n_2 \}$ describing an event $E$ at time $t$ with weight $w$. This list of vectors defines color dipoles with which we associate a virtual correction
\begin{align}
    V_E = V_{1i_1}+V_{i_1i_2} + \cdots + V_{i_{m-2}2}\, ,
\end{align}
where we used
\begin{align}
    V_{ij} = 4 N_c\int [\mathrm{d}^2\Omega_l]  \spac W_{ij}^l \,.
\end{align}
The time steps in our LL shower are computed according to the distribution
\begin{equation}
P_E(\Delta t) = V_E\, e^{-V_E \Delta t}\,,
\end{equation}
after which an additional parton is generated and added to the event. The shower terminates if the additional parton is in the veto region, see Figure \ref{fig:LLshower}.

To compute the correction \eqref{eq:2looprunningApp}, we run two showers. At shower time $t_0=0$ we start a LL shower with a list of vectors $\{\underline{n}\}$ along the direction of the hard partons in the Born level process and weight $w=1/V_E$. In our case $\{\underline{n}\}=\{n_1,n_2\}$ contains two particles which are back-to-back. The purpose of the first shower is to generate event configurations with different amounts of particles at subsequent times. For instance, on the left of Figure \ref{fig:LLshower} we have
\begin{equation}
\begin{aligned}
    t_1: \hspace{1mm} & E_1 = \{n_1,n_2\}\,,\\
    t_2: \hspace{1mm} & E_2 = \{n_1,n_3,n_2\}\,,\\
    t_3: \hspace{1mm} & E_3 = \{n_1,n_3,n_4,n_2\}\,,\\
    & \hspace{6.2mm} \vdots \\
    t_8: \hspace{1mm} & E_{8} = \{n_1,n_6,n_9,n_3,n_8,n_5,n_4,n_7,n_2\}\,.
\end{aligned}
\end{equation}
This represents $\bm{U}_{2k'}(t_0,t')$ in \eqref{eq:2looprunningApp}. At $t^\prime=t_1,\spac t_2,\spac \dots, $ we then insert the two-loop anomalous dimension correction, which increases the number of vectors by $0$, $1$ or $2$, depending on whether $\bm{v}_m$, $\bm{r}_m$ or $\bm{d}_m$ is computed. After this, we start a second LL shower, which corresponds to the factor $\bm{U}_{l'l}(t',t)$ in \eqref{eq:2looprunningApp}. This is depicted on the right-hand side of Figure \ref{fig:LLshower} for the example of $t'=t_2$.

Let us now explain in detail how the insertion is implemented in our shower code, using, as an example, a term involving a single emission, such as the $X_{ij}^q$ term in $\bm{r}_m$, or the single angular integral terms in $\bm{v}_m$.
\begin{enumerate}
    \item Pick a dipole in $E_i$ with the probability $V_{ij} / V_{E_i}$. Then generate an additional direction $n_q$ and evaluate its weight $\Delta \Gamma$, which is just the integrand of one of the terms in $\bm{\Gamma}^{(2)}-{\beta_1}/{\beta_0} \spac \bm{\Gamma}^{(1)}$ under consideration. 
    
\item  For real contributions we need to update the list of vectors by inserting the additional emission between its parents, i.e. $E_i' = \{n_1, \dots,n_i,n_q,n_j,\cdots,n_2 \}$, while we leave $E_i'= E_i$ for virtual contributions.

    \item Compute an insertion weight
    \begin{align}\label{eq:insertionweight}
         w_I = \Delta \Gamma\, \frac{V_{E_i}}{V_{ij}} \, ,
    \end{align}
    where the factor $V_{E_i}/V_{ij}$ cancels the one introduced when selecting the dipole.
    \item Start a LL shower with weight $w_{\rm new} = w_I/V_{E_i'} $ and list of vectors $E_i'$ at $t'$. The time values arising in the second shower correspond to values of $t$ in \eqref{eq:2looprunningApp} and the weight $w_{\rm new}$ is filled into a histogram at each time $t$. 
\end{enumerate}
For concreteness, we described the implementation of terms involving a single emission or a single angular integral. The contributions with two directions $n_q,n_r$ are implemented analogously. By repeating the entire procedure $N$ times and averaging the histograms, we get a numerical estimate for the integral in \eqref{eq:2looprunningApp}. 

\subsection{D. Comparison with finite-$N_c$ calculation at LL}
\renewcommand{\theequation}{D.\arabic{equation}}
\setcounter{equation}{0}
\begin{figure}
\includegraphics[scale=0.5]{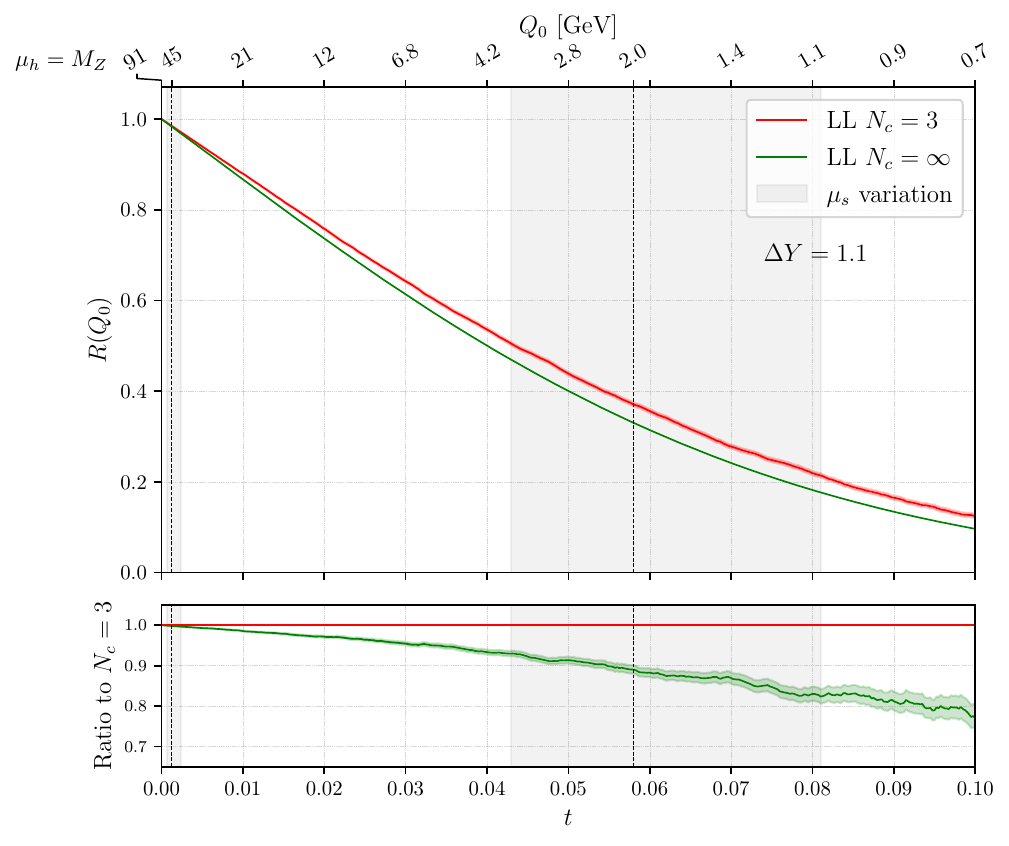}\hspace{10mm}
\includegraphics[scale=0.5]{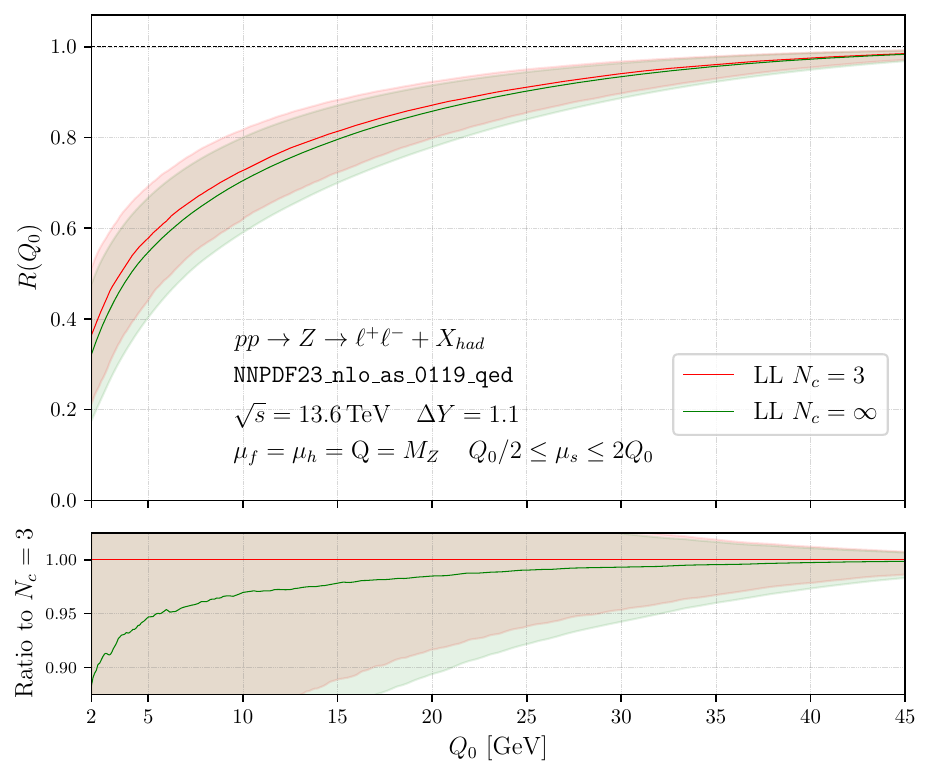}
\caption{\label{fig:LL_hatta_vs_nico_3} Comparison between the $N_c=3$ (red) and the large-$N_c$ (green) results for $Z$-boson prediction at LL accuracy.}
\end{figure}

In \cite{Hatta:2013iba} the interjet energy flow at LL was computed for $N_c=3$. We have included this result in the plots presented in the main text. To illustrate the size of the finite-$N_c$ corrections to the interjet energy flow, we now compare to the LL result obtained using the parton shower in the large-$N_c$ limit. In the left panel of Figure \ref{fig:LL_hatta_vs_nico_3} we show the LL results as a function of the evolution time $t$. The gap fraction is normalized to one at $t=0$ and the deviations between the two results then become larger with increasing $t$. On the plot's upper edge, we show the corresponding $Q_0$ values for $\mu_h=M_Z$ for illustrative purposes. The black dashed lines represent the values $Q_0 =45\,{\rm GeV}$ (left line) and $Q_0=2\,{\rm GeV}$ (right line). We see that the finite-$N_c$ corrections within the energy range we consider in the main text reach at most $10\%$.

A variation of the scale $\mu_s$ translates into a variation of the value of $t$ and the gray shaded areas in the left plot show the $\mu_s$ scale variations around $2\,{\rm GeV}$ and $45\,{\rm GeV}$. We compute the scale from the profile function $\mu_s(x \spac Q_0,\mu_h)$ introduced in \cite{Balsiger:2019tne} and vary $1/2 < x < 2$. The width of the band becomes smaller at larger $Q_0$ as the profile function switches off resummation and enforces $\mu_s(x\spac Q_0,\mu_h)\to \mu_h$ for $Q_0\to\mu_h$. In the right panel of Figure \ref{fig:LL_hatta_vs_nico_3}, we show the same plot as a function of $Q_0$. We see that the scale uncertainty of the LL result by itself is larger than the finite-$N_c$ effects so their inclusion does not improve the accuracy of the predictions unless NLL corrections are included as well. 

\subsection{E. Consistency checks and comparison with \textsc{Gnole}}
\renewcommand{\theequation}{E.\arabic{equation}}
\setcounter{equation}{0}
In \cite{Banfi:2021xzn} first results for the full set of NLL corrections to the interjet energy flow were obtained using the computer code \textsc{Gnole} which is based on the framework presented in \cite{Banfi:2021owj}. Our LL results agree within statistical fluctuations to few permille accuracy with \textsc{Gnole} over the range $t<0.06$ relevant for our predictions. This is not surprising since a previous version of our code \cite{Balsiger:2018ezi} has been cross checked against the results of \cite{Dasgupta:2001sh}. 

The matching corrections $\bm{\mathcal{H}}_2^{(1)}$ and $\bm{\mathcal{H}}_3^{(1)}$, required at subleading accuracy, agree as well. These matching corrections are calculated using the LL shower and do not probe contributions due to the two-loop anomalous dimension. To test the latter, we isolate contributions in \eqref{eq:gamma2} proportional to $\gamma_{\rm cusp}$. This is the simplest part of $\bm{\Gamma}^{(2)}$, and we use it to perform a series of consistency checks to validate that our algorithm correctly implements \eqref{eq:2looprunningApp}. First, we can exponentiate the $\gamma_{\rm cusp}$ contributions and subsequently expand numerically. This is delicate since we have to numerically take the limit of $\alpha_s \rightarrow 0$ in order to suppress NNLL terms. After taking the limit we can compare to the result with one insertion of the anomalous dimension. As expected, we find agreement within our framework. 
In addition, we compare to the $\gamma_{\rm cusp}$-piece computed with \textsc{Gnole} and agree to better than a percent for $t<0.06$.

It is more difficult to test the remaining parts of \eqref{eq:gamma2} which involve the angular functions $K_{ij;qr}$ and $M_{ij;qr}$. One important test is that our numerical code reproduces the analytical integrals over these expressions which arise in the NNLO correction of the interjet energy flow at small $Q_0$ given in \cite{Becher:2021urs}, for example for the combination
\begin{align}
    \int \big[d^2\Omega_r\big]\int \big[d^2\Omega_q\big]\big(K_{12;qr}+K_{21;qr}\big)\spac \theta_{\mathrm{in}}(n_q)\spac \spac \theta_{\mathrm{out}}(n_q)\,.
\end{align}
We have calculated this quantity with our shower and find that the integration converges to the analytical result. Together with the validation of our implementation of the single insertion of $\bm{\Gamma}^{(2)}$ using the $\gamma_{\rm cusp}$-piece, this provides a strong check on our shower code.

It is not obvious how the more complicated parts of $\bm{\Gamma}^{(2)}$ map between \textsc{Gnole} and our RG framework. The \textsc{Gnole} approach works with full four-vectors instead of directions and to obtain the mapping analytically one would have to integrate out the energies of the generated partons. It will be interesting to map the two formalisms analytically, but for the moment we perform the comparison numerically. We find agreement with {\sc Gnole} within numerical uncertainties, at the level of a few per cent.\footnote{We thank Pier Monni for modifying his code to simplify the comparison and for providing his {\sc Gnole} results.}

\end{document}